\documentclass[aps,twocolumn,showpacs]{revtex4}

\usepackage{graphicx}
\usepackage{epstopdf}
\usepackage{dcolumn}
\usepackage{bm}
\usepackage{amsmath}
\usepackage{amssymb}
\usepackage{array}
\usepackage[caption=false]{subfig}
\usepackage[bookmarks=false,linkcolor=blue,urlcolor=blue,colorlinks,citecolor=blue]{hyperref}
\usepackage{exscale,relsize}
\usepackage{lipsum}
\usepackage{appendix}

\def \eps{{\varepsilon}}
\def \om{{\omega}}
\def \ua{{\uparrow}}
\def \da{{\downarrow}}
\def \bs#1{{\boldsymbol#1}}
\def \tx#1{{\rm#1}}
\def \mc#1{\mathcal{#1}}
\def \avg#1{\langle #1 \rangle}

\DeclareMathOperator{\sgn}{sgn}

\begin{document}

\title{Interaction-driven topological superconductivity in one dimension}
\author{Arbel Haim$^1$, Konrad W\"olms$^2$, Erez Berg$^1$, Yuval Oreg$^1$, and Karsten Flensberg$^2$}
\affiliation{$^1$Department of Condensed Matter Physics$,$ Weizmann Institute of Science$,$ Rehovot$,$ 76100$,$ Israel\\
\mbox{$^2$Center for Quantum Devices, Niels Bohr Institute$,$ University of Copenhagen$,$ DK-2100 Copenhagen \O, Denmark}}
\date{\today}

\begin{abstract}
We study one-dimensional topological superconductivity in the presence of time-reversal symmetry. This phase is characterized by having a bulk gap, while supporting a Kramers' pair of zero-energy Majorana bound states at each of its ends. We present a general simple model which is driven into this topological phase in the presence of repulsive electron-electron interactions. We further propose two experimental setups and show that they realize this model at low energies. The first setup is a narrow two-dimensional topological insulator partially covered by a conventional $s$-wave superconductor, and the second is a semiconductor wire in proximity to an $s$-wave superconductor. These systems can therefore be used to realize and probe the time-reversal invariant topological superconducting phase. The effect of interactions is studied using both a mean-field approach and a renormalization group analysis.
\end{abstract}

\maketitle

\section{Introduction}
\label{sec:intro}

The pursuit of realizing topological phases in condensed matter systems continues. These phases are generally characterized by having unique surface properties which are dictated by the topological properties of the bulk. The first and most famous example is the quantum Hall effect (QHE)~\cite{Klitzing1980new,Laughlin1981quantized,Thouless1982quantized}, in which gapless chiral edge modes, protected only by topology, reside on the edges of a two-dimensional system and give rise to a quantized Hall conductivity.

Since then it has been realized that upon invoking symmetries, a rich variety of topological phases can emerge~\cite{schnyder2008classification,kitaev2009periodic,Qi2011topological}. These phases as well contain gapless boundary modes which are related to the topological nature of the bulk. However, they are only protected in the presence of some imposed symmetries, and could otherwise become gapped. Here, the paradigmatic example is the topological insulator (TI)~\cite{Kane2005quantum,Bernevig2006quantum,konig2007quantum} which in two dimensions can be thought of as two copies of the QHE, related by time-reversal transformation. The edge of the system now host gapless \emph{helical} modes which are protected by the presence of time-reversal symmetry (TRS).

The various topological phases are classified according to the possible symmetries present in a given system~\cite{schnyder2008classification,kitaev2009periodic}. These are TRS, particle-hole symmetry (PHS) and chiral symmetry~\cite{Altland1997}. Of particular interest is the so-called class-D topological superconductor (TSC)~\cite{Alicea2012,Beenakker2013} which is protected solely by PHS. This symmetry is special since it exists in all superconducting systems, and in fact cannot truly be broken. This makes its edge states, the Majorana modes, extremely robust. In that sense, the TSC can be viewed as the superconducting analog of the QHE.

One is then prompted to ask: what is the superconducting analog of the topological insulator? This would be the time-reversal invariant topological superconductor (TRITOPS) which belongs to class DIII~\cite{Qi2009time,Qi2010topological}. In one or two dimensions, it can be described as two copies of a class D TSC, related by time-reversal transformation. Each edge (or end) of this phase hosts a Kramers' pair of time-reversal related Majorana modes, analogous to the pair of helical edge modes of the two-dimensional (2d) TI. Unlike single Majorana zero modes, Majorana Kramers pairs do not have a well defined braiding statistics~\cite{Wolms2014local,Wolms2016Braiding}; however, they have non-trivial spin structure~\cite{Keselman2013inducing,Keselman2015gapless}.

Class-D TSC is currently a subject of intense study, following the prediction that this phase can be engineered by combining well-understood building blocks such as conventional $s$-wave superconductors and spin orbit-coupled material~\cite{fu2008superconducting,fu2009josephson,Lutchyn2010majorana,Oreg2010helical}. Recently a number of experimental studies have shown evidence consistent with the existence of zero-energy Majorana bound state (MBS) in such one-dimensional (1d) systems~\cite{mourik2012signatures,deng2012anomalous,Das2012zero,churchill2013superconductor,Finck2013anomalous,Nadj-Perge2014observation,Pawlak2015probing,Ruby2015end}.

In contrast, to the best of our knowledge, no attempts have been made on experimentally realizing the TRITOPS phase in low-dimensional systems. Some theoretical works have proposed using unconventional superconductivity in order to realize this phase~\cite{Wong2012majorana,Zhang2013time,Nakosai2013majorana}. Other proposals include $\pi$ junctions~\cite{Dahlhaus2010Random,Keselman2013inducing,Schrade2015proximity}, organic SCs~\cite{Dumitrescu2013topological}, and intrinsic superconductors in two and three dimensions~\cite{Fu2010,Nakosai2012topological,Deng2012majorana,Wang2014two}. In particular, it has been shown~\cite{Zhang2013time,Gaidamauskas2014majorana,Haim2016WithComment} that, unlike class-D TSC, the TRITOPS phase cannot be engineered by proximity coupling a conventional $s$-wave superconductor (SC) to a system of noninteracting electrons.

It was suggested~\cite{Gaidamauskas2014majorana,Haim2014time,Klinovaja2014time,Klinovaja2014Kramers} that repulsive interactions in a 1d system proximity-coupled to a convention $s$-wave SC can stabilize the TRITOPS phase. This mechanism was demonstrated explicitly in a proximity coupled semiconductor nanowire using a mean-field approximation~\cite{Haim2014time,Danon2015interaction} and using the density matrix renormalization group~\cite{Haim2014time}. It has also been suggested that interactions can induce (gapless) topological phases supporting Majorana zero-modes in 1d, with~\cite{Cho2014topological,Keselman2015gapless,Kainaris2015Emergent} and without~\cite{Sau2011Number,Fidkowski2011Majorana,Ruhman2015Topological} time-reversal symmetry, even in the absence of proximity to a bulk SC.

In this paper we adopt a more general perspective of interaction-driven TRITOPS. We consider a general ``minimal'' model (see Fig.~\ref{fig:H_0_spectrum}) which can arise as a low-energy theory of various spin-orbit coupled 1d systems in proximity to an $s$-wave SC. The model has four Fermi points with two right moving modes and two left-moving modes. Due to spin-orbit coupling, proximity-induced superconductivity results in both a singlet and a triplet pairing potential, $\Delta_\tx{s}$ and $\Delta_\tx{t}$, respectively~\cite{Gorkov2001WithComment}. As we show, short-range repulsive interactions suppress $\Delta_\tx{s}$ compared to $\Delta_\tx{t}$~\cite{Sun2014tuningWithComment}, thereby driving the system into the topological phase. We map the phase diagram of this minimal model using both a mean-field approximation and an analytically controlled renormalization group (RG) analysis. We further propose two microscopic systems which can be realized in currently-available experimental setups, and which are described at low energies by the minimal model. These are (i) a narrow 2d TI partially covered by an $s$-wave SC [see Fig.~\hyperref[fig:Narrow_QSHI]{\ref{fig:Narrow_QSHI}(a)}], and (ii) a quasi 1d semiconductor nanowire proximitized by an $s$-wave SC [see Fig.~\hyperref[fig:wire_dispersion]{\ref{fig:wire_dispersion}(a)}].

While we consider clean systems in this work, we expect our results to hold also for systems with weak disorder. Namely, we expect the topological phase to survive as long as the mean free time associated with disorder is large compared to the inverse energy gap, similar to the case of the class D topological SC~\cite{Motrunich2001Griffiths,Brouwer2011Probability}.

Several studies have examined the effect of repulsive interactions on topological superconductors with \emph{broken} time-reversal symmetry~\cite{Gangadharaiah2011Majorana,Stoudenmire2011Interaction,Manolescu2014Coulomb}. It was found that the topological phase is stable against moderate interactions which do not close the bulk energy gap. In this paper, on the other hand, we are interested in the time-reversal symmetric phase. Importantly, while in the above studies the topological phase exists even in the absence of interactions, here the role of repulsive interactions is a crucial one; they are responsible for driving the system into the topological phase.

The rest of this paper is organized as follows. In Sec.~\ref{sec:The_model} we introduce the low-energy minimal model and the conditions for it to be in its non-trivial phase. In Sec.~\ref{sec:Realizations} we examine the two microscopic models mentioned above and show that they are described at low energies by the minimal model. We then study the effect of repulsive interactions in Sec.~\ref{sec:interactions}, showing that it drives the low-energy model into the topological phase. This is done first on a mean-field level in Sec.~\ref{sec:MF}, and then using a perturbative RG analysis in Sec.~\ref{sec:RG}. We conclude and discuss our results in Sec.~\ref{sec:discussion}.

\section{Main Theme}
\label{sec:The_model}

\begin{figure}
\includegraphics[clip=true,trim =0mm 0mm 0mm 0mm,width=0.4\textwidth]{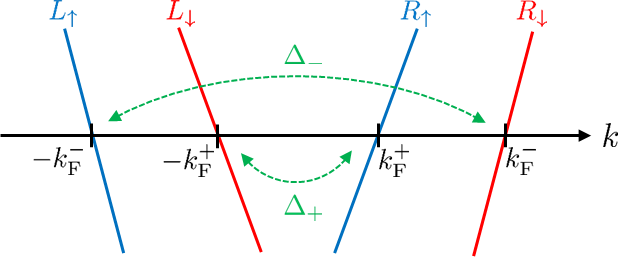}
\caption{Dispersion of the low-energy Hamiltonian $H_0$, having two right-moving modes and two left-moving modes [see Eq.~\eqref{eq:minimal_H}]. The Hamiltonian $H_\Delta$ describes induced superconductivity. The pairing potential $\Delta_+$ couples the modes of positive helicity, while $\Delta_-$ couples the modes of negative helicity. The system is in its topologically nontrivial phase when $\sgn(\Delta_+)\sgn(\Delta_-)=-1$ [see Eq.~\eqref{eq:top_inv}].}\label{fig:H_0_spectrum}
\end{figure}

Our minimal model is described in the absence of interactions by the following Hamiltonian
\begin{equation}\label{eq:minimal_H}
\begin{split}
&H = H_0+H_\Delta,\\
&H_0 = -i\int \tx{d}x
\left\{ v_+  \left[R^\dag_\ua(x)\partial_x R_\ua(x) -L^\dag_\da(x)\partial_x L_\da(x)\right]\right.\\
& \hskip 20mm
+ v_- \left.  \left[R^\dag_\da(x)\partial_x R_\da(x) -L^\dag_\ua(x)\partial_x L_\ua(x)\right]\right\},\\
&H_\Delta = \int \tx{d}x \left[\Delta_+ R^\dag_\ua(x)L^\dag_\da(x) + \Delta_- L^\dag_\ua(x)R^\dag_\da(x) + \tx{h.c.} \right],
\end{split}
\end{equation}
where $R_s$ ($L_s$) is an annihilation operator of a right (left) moving fermionic mode of spin $s$. Here, $\Delta_+$ and $\Delta_-$ are two induced pairing potentials. $\Delta_+$ describes pairing between the modes of positive helicity, $R_\uparrow$ and $L_\downarrow$, while $\Delta_-$ describes pairing between the modes of negative helicity, $L_\uparrow$ and $R_\downarrow$~\cite{PseudoSpinIndex}. Similarly, $v_\pm$ are the velocities of the modes with positive and negative helicity, respectively. The dispersion of $H_0$ is shown in Fig.~\ref{fig:H_0_spectrum}.

The time-reversal operation is defined by
\begin{equation}\label{eq:TR_operation}
\begin{split}
&\mathbb{T}R_s(x)\mathbb{T}^{-1} = i\sigma^y_{ss'} L_{s'}(x)\\
&\mathbb{T}L_s(x)\mathbb{T}^{-1} = i\sigma^y_{ss'} R_{s'}(x)\\
&\mathbb{T}i\mathbb{T}^{-1}= -i,
\end{split}
\end{equation}
where $\{\sigma^i\}_{i=x,y,z}$ is the set of Pauli matrices operating in spin space. Requiring that $H$ obeys time-reversal symmetry, $\mathbb{T}H\mathbb{T}^{-1}=H$, imposes the constraints that both $\Delta_+$ and $\Delta_-$ are real. In the absence of inversion symmetry, the Fermi momenta $k_\tx{F}^+$ and $k_\tx{F}^-$ generally differ from one another (see Fig.~\ref{fig:H_0_spectrum}). In this case, $H$ is the most general low-energy quadratic Hamiltonian which describes a single-channel 1d system with TRS~\cite{MomentumMismachTerm}.

The time-reversal operation in Eq.~\eqref{eq:TR_operation} squares to $-1$, placing this system in class DIII of the Altland-Zirnbauer classification~\cite{Altland1997}, with a $\mathbb{Z}_2$ topological invariant~\cite{schnyder2008classification,kitaev2009periodic}. It has been shown by Qi \emph{et al.}~\cite{Qi2010topological} that the topological invariant of this class in 1d is determined by the product of the signs of the pairing potentials at the Fermi points~\cite{TopInvWeaPairing}. Applying their result to our case reads
\begin{equation}\label{eq:top_inv}
\mathcal{Q}=\sgn(\Delta_+)\cdot\sgn(\Delta_-),
\end{equation}
where $Q=-1$ corresponds to the topologically non-trivial phase, having a Kramers' pair of Majorana bound states at each end of the system. For completeness we derive this result in Appendix~\ref{sec:top_criter} using a scattering-matrix approach.

It is instructive to write the superconducting part of the Hamiltonian in the following form
\begin{equation}\label{eq:H_delta_2}
\begin{split}
H_\Delta =& \int \tx{d}x \left\{\Delta_\tx{s}\left[R^\dag_\ua(x)L^\dag_\da(x) - R^\dag_\da(x)L^\dag_\ua(x)\right]\right.\\
+&\left.\Delta_\tx{t}\left[R^\dag_\ua(x)L^\dag_\da(x) + R^\dag_\da(x)L^\dag_\ua(x)\right] +\tx{h.c.} \right\},
\end{split}
\end{equation}
where $\Delta_\tx{s,t}=(\Delta_+ \pm \Delta_-)/2$ are the singlet and triplet pairing potentials respectively. Inserting this in Eq.~\eqref{eq:top_inv} results in
\begin{equation}\label{eq:top_inv_2}
\mathcal{Q}=\sgn(\Delta_\tx{s}^2-\Delta_\tx{t}^2).
\end{equation}
Namely, the topological phase ($\mathcal{Q}=-1$) is obtained when the triplet pairing term exceeds in magnitude the singlet pairing term.

For a noninteracting system in proximity to a conventional $s$-wave SC the system will always be in the topologically trivial phase~\cite{Zhang2013time,Gaidamauskas2014majorana,Haim2016WithComment}, namely $|\Delta_\tx{s}|\ge |\Delta_\tx{t}|$. In Secs.~\ref{sec:MF} and~\ref{sec:RG} we will show that repulsive short-range interactions effectively suppress the singlet pairing term $\Delta_\tx{s}$ in comparison with the triplet term $\Delta_\tx{t}$. Depending on the bare ratio $|\Delta_\tx{t}|/|\Delta_\tx{s}|$, strong enough interactions can therefore drive the system to the topological phase~\cite{IntSysTopInvComment}.
A system in which initially $|\Delta_\tx{t}|$ is of the order of (but less than) $|\Delta_\tx{s}|$, is therefore more susceptible to become topological by the presence of repulsive interactions.

Before studying in detail the effect of repulsive interactions in the proximitized system, we first present two examples of microscopic models for systems which are described at low energies by the Hamiltonian of Eq.~\eqref{eq:minimal_H}. Importantly, we show that the low-energy Hamiltonian for these systems contains a bare nonvanishing induced triplet term which is generally of the order of (but smaller than) the singlet term.

\section{Realizations}
\label{sec:Realizations}

In this section we concentrate on two specific examples of proximity-coupled systems where both a singlet and a triplet pairing terms are induced. We then move on to show in Sec.~\ref{sec:interactions} that repulsive electron-electron interactions suppress the singlet pairing compared to the triplet pairing, thereby driving the systems to the TRITOPS phase.

\subsection{Narrow Quantum Spin-Hall Insulator}
\label{sec:Narrow_QSHI}

\begin{figure}
\begin{tabular}{lrr}
\rlap{\parbox[c]{0.0cm}{\vspace{-4.7cm}\footnotesize{(a)}}}
\hskip -2mm
\includegraphics[clip=true,trim =0cm -0.4cm 0cm 0.0cm,width=0.166\textwidth]{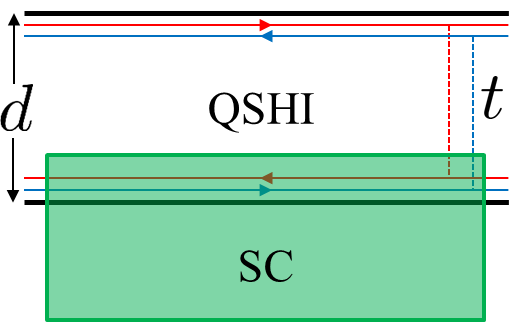} & \hskip 1mm
\rlap{\parbox[c]{0.0cm}{\vspace{-4.7cm}\footnotesize{(b)}}}
\includegraphics[clip=true,trim =0cm 0cm 0cm 0cm,width=0.168\textwidth]{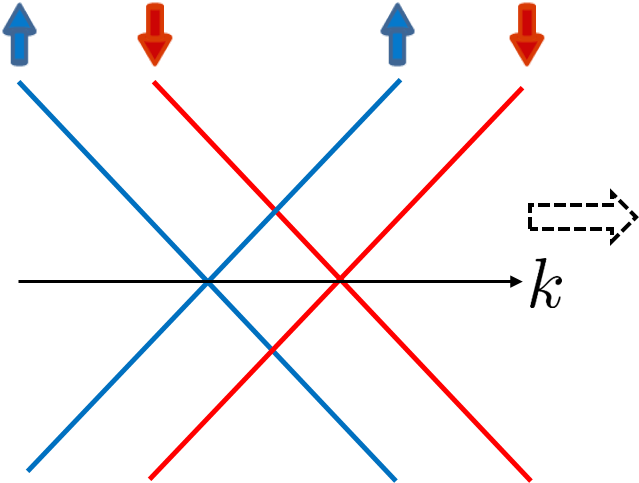} & \hskip -4.2mm
\rlap{\parbox[c]{0.0cm}{\vspace{-4.7cm}\footnotesize{(c)}}}
\includegraphics[clip=true,trim =0cm 0cm 0cm 0cm,width=0.146\textwidth]{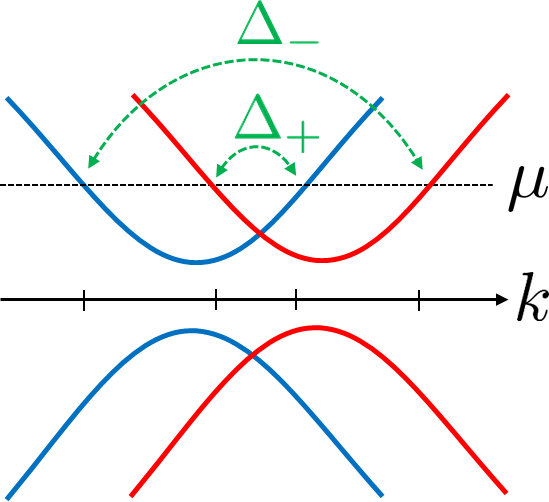}
\end{tabular}
\caption{(a) A narrow two dimensional quantum spin Hall insulator in proximity to a conventional $s$-wave superconductor. (b) In the absence of induced superconductivity and tunneling between opposite edges, the low-energy electronic spectrum is described by two pairs of helical edge states, the helicity being opposite for the two edges. (c) If the width of the sample $d$ is of the order of the characteristic correlation length $\xi_\tx{QSHI}$ or less, then the opposite edges are coupled (with a coupling constant $t$), and an energy gap is opened. Such a coupling between the edge modes is necessary in order to have nonzero backscattering interaction, which is crucial for realizing the topological phase (see Sec.~\ref{sec:interactions}).}\label{fig:Narrow_QSHI}
\end{figure}

We consider a narrow two-dimensional quantum spin Hall insulator (QSHI) in proximity to an $s$-wave SC~\cite{Hart2014Induced}. A QSHI~\cite{Kane2005quantum,Bernevig2006quantum,konig2007quantum} is a phase characterized by a pair of counter-propagating helical modes on each edge of the system as depicted in Fig.~\hyperref[fig:Narrow_QSHI]{\ref{fig:Narrow_QSHI}(a)}. We define the correlation length $\xi_\tx{QSHI}$ as the characteristic length with which the helical edge modes decay into the bulk. If the width of the bar $d$ is of the order of $\xi_\tx{QSHI}$ or less, then gapless modes of opposite edges are coupled and an energy gap is opened [cf. Figs.~\hyperref[fig:Narrow_QSHI]{\ref{fig:Narrow_QSHI}(b)} and~\hyperref[fig:Narrow_QSHI]{\ref{fig:Narrow_QSHI}(c)}]. When the chemical potential lies above or below the gap, the low-energy sector of the system is described by a one-dimensional model having four Fermi points, similar to the Hamiltonian $H_0$ of Eq.~\eqref{eq:minimal_H}. We now show that coupling one of the edges to a conventional $s$-wave SC results in a nonvanishing triplet pairing component.

In the absence of interactions, the two coupled edges are described by the following Hamiltonian
\begin{equation}
\begin{split}
&H_\tx{QSHI}=\sum_k \Psi_k^\dag\mathcal{H}(k)\Psi_k^{\phantom{\dag}}\hskip 2mm ;\hskip 2mm
\Psi^\dag_k=(a^\dag_{k\uparrow},b^\dag_{k\uparrow},a^{\phantom{\dag}}_{-k\downarrow},b^{\phantom{\dag}}_{-k\downarrow}) \\
&\mathcal{H}(k)=[-\mu+(\delta\mu +vk)\lambda^z+t\lambda^x]\tau^z+\frac{\Delta_{\rm ind}}{2}(1+\lambda^z)\tau^x,
\end{split}
\end{equation}
where $a^\dag_{ks}$ ($b^\dag_{ks}$) creates an electron with momentum $k$ and spin $s=\uparrow,\downarrow$ on the lower (upper) edge of the sample. $\{\tau^i\}_{i=x,y,z}$ and $\{\lambda^i\}_{i=x,y,z}$ are sets of Pauli matrices in the particle-hole space and the lower edge-upper edge space, respectively. Here, $v$ is the propagation velocity of the edge modes, $t$ is the coupling constant between the lower and upper edge modes (which results from the finite width of the sample), $\mu\pm\delta\mu$ are the chemical potentials at the upper and lower edge, respectively, and $\Delta_{\rm ind}$ is the pairing potential induced by the SC on the lower edge of the sample [cf. Fig.\hyperref[fig:Narrow_QSHI]{~\ref{fig:Narrow_QSHI}(a)}].

We consider the case where, in the absence of proximity, the chemical potential lies inside the upper band [see Fig.~\hyperref[fig:Narrow_QSHI]{\ref{fig:Narrow_QSHI}(c)}]~\cite{UpperVsLowerBand}, and where the induced pairing, $\Delta_\tx{ind}$, is small in comparison with the distance to the lower band, $\mu+|t|$. We can therefore project out the lower band, arriving at the following effective Hamiltonian for the upper band
\begin{equation}
\begin{split}
&H_{\rm eff}=\sum_k\left\{
\sum_{s=\ua,\da}\left(\sqrt{t^2+(\delta\mu+ svk)^2}-\mu\right)c^\dag_{ks}c^{\phantom{\dag}}_{ks}\right. \\
&\left.\phantom{\sum_{s=\ua,\da}}\hskip 10mm+\Delta(k)\left(c^\dag_{k\uparrow}c^\dag_{-k\downarrow}+{\rm h.c.}\right)\right\},
\end{split}\label{eq:H_eff_QSHI}
\end{equation}
with the effective pairing potential
\begin{equation}
\Delta(k)=\frac{\Delta_\tx{ind}}{2}\left[1+(\delta\mu+vk)/\sqrt{t^2+(\delta\mu+vk)^2}\right].
\end{equation}
Here, $c^\dag_{ks}$ describes electronic modes in the upper band with momentum $k$ and spin $s$. It is related to the left and right edge modes through
\begin{equation}\label{eq:qshi_modes}
\begin{split}
  &c^\dag_{ks}=\cos(\phi_{ks}) a^\dag_{ks} + \sin(\phi_{ks})b^\dag_{ks},\\
  &\cos(2\phi_{ks})=(\delta\mu+vks)/\sqrt{t^2+(\delta\mu+vks)^2},\\
  &\sin(2\phi_{ks})=t/\sqrt{t^2+(\delta\mu+vks)^2},
\end{split}
\end{equation}
where we have used a convention in which $s=1$ corresponds to spin $\uparrow$, and $s=-1$ corresponds to spin $\downarrow$.

Assuming weak pairing~\cite{WeakPairing}, we can linearize the spectrum near the Fermi energy and impose a momentum cutoff $\Lambda$, resulting in the following Hamiltonian
\begin{equation}\label{eq:H_lin}
\begin{split}
H^\tx{lin} = \sum_{|k|<\Lambda} &\Big\{ \bar{v}k\sum_{s=\ua,\da} (R^\dag_{ks}R^{\phantom{\dag}}_{ks}-L^\dag_{ks}L^{\phantom{\dag}}_{ks}) \\
+& ( \Delta_+ R^\dag_{k\uparrow}L^\dag_{-k\downarrow}
+\Delta_-L^\dag_{k\uparrow}R^\dag_{-k\downarrow}+{\rm h.c.} ) \Big\}
\end{split}
\end{equation}
where
\begin{equation}\label{eq:def_chirl_fields_qshi}
\begin{split}
R_{k\uparrow} &= c_{k^{+}_{\rm F}+k,\uparrow} \hskip 5mm ; \hskip 5mm
L_{k\downarrow} = c_{-k^{+}_{\rm F}+k,\downarrow},\\
R_{k\downarrow} &= c_{k^{-}_{\rm F}+k,\downarrow} \hskip 5mm ; \hskip 5mm
L_{k\uparrow} = c_{-k^{-}_{\rm F}+k,\uparrow},
\end{split}
\end{equation}
and
\begin{equation}
  \Delta_+=\Delta(k_{\rm F}^{+}) \hskip 3mm , \hskip 3mm \Delta_-=\Delta(-k_{\rm F}^{-}).
\end{equation}
The velocity of the modes at the Fermi points is given by
\begin{equation}\label{eq:velocity_qshi}
\bar{v}=v\sqrt{1-(t/\mu)^2},
\end{equation}
and the Fermi momenta are given by $k^{\pm}_\tx{F}=(\mu \bar{v}/v \mp \delta\mu)/v$ (Notice that since the chemical potential is assumed to lie inside the upper band one has $\mu>|t|$). The Hamiltonian of Eq.~\eqref{eq:H_lin} is exactly the minimal model Hamiltonian of Eq.~\eqref{eq:minimal_H}, written in momentum space, with $v_+=v_-=\bar{v}$.

As discussed in Sec.~\ref{sec:The_model}, the system is in its topological phase when $\sgn(\Delta_+)\sgn(\Delta_-)=-1$. Alternatively stated, this occurs when $|\Delta_{\rm t}|>|\Delta_{\rm s}|$ [see Eq.~\eqref{eq:H_delta_2}]. For the model at hand one has
\begin{equation}\label{eq:t_vs_s_QSHI}
  \Delta_\tx{s}=\Delta_\tx{ind}/2 \hskip 3mm ; \hskip 3mm \Delta_\tx{t}= \frac{\bar{v}}{v} \Delta_\tx{ind}/2.
\end{equation}
As expected, in the absence of interactions $|\Delta_\tx{t}|\le|\Delta_\tx{s}|$. Importantly, however, $\Delta_\tx{t}$ is nonzero, and can generally be of similar magnitude to $\Delta_\tx{s}$, making the system susceptible to being driven into the TRITOPS phase by short-range repulsive interactions.

The existence of a nonvanishing triplet pairing term can also be understood from a simple qualitative argument. The lower and upper edges of the QSHI host modes of positive and negative helicity, respectively. Since the SC is coupled to the \emph{lower} edge, the pairing of the positive-helicity modes, $\Delta_+$, is larger in magnitude than that of the negative-helicity modes, $\Delta_-$, and consequently $\Delta_\tx{t}\neq 0$. This agrees with Eqs.~\eqref{eq:velocity_qshi} and~\eqref{eq:t_vs_s_QSHI} which suggest that $|\Delta_\tx{t}|$ is maximal when the edges are maximally separated (namely when $t=0$). We note, however, that some overlap between the edge modes is necessary in order to eventually achieve the TRITOPS phase. This is because in the absence of such overlap, the backscattering interaction vanishes. As will be shown in Sec.~\ref{sec:interactions}, this interaction terms is crucial for the system to be driven into the topological phase.

\subsection{Proximity-Coupled Semiconductor Wire}
\label{sec:Rashba_wire}

Next we concentrate on another system which can be driven into the TRITOPS phase by repulsive interactions, a spin-orbit coupled semiconductor nanowire. We now show that this system is described at low-energies by the Hamiltonian of Eq.~\eqref{eq:minimal_H} with a nonvanishing triplet pairing term.

\begin{figure}
\begin{tabular}{lr}
\rlap{\parbox[c]{7.5cm}{\vspace{-4.5cm}\footnotesize{(a)}}}
\hskip -2mm
\includegraphics[clip=true,trim =0cm 0.5cm 0cm 0cm,scale=0.23]{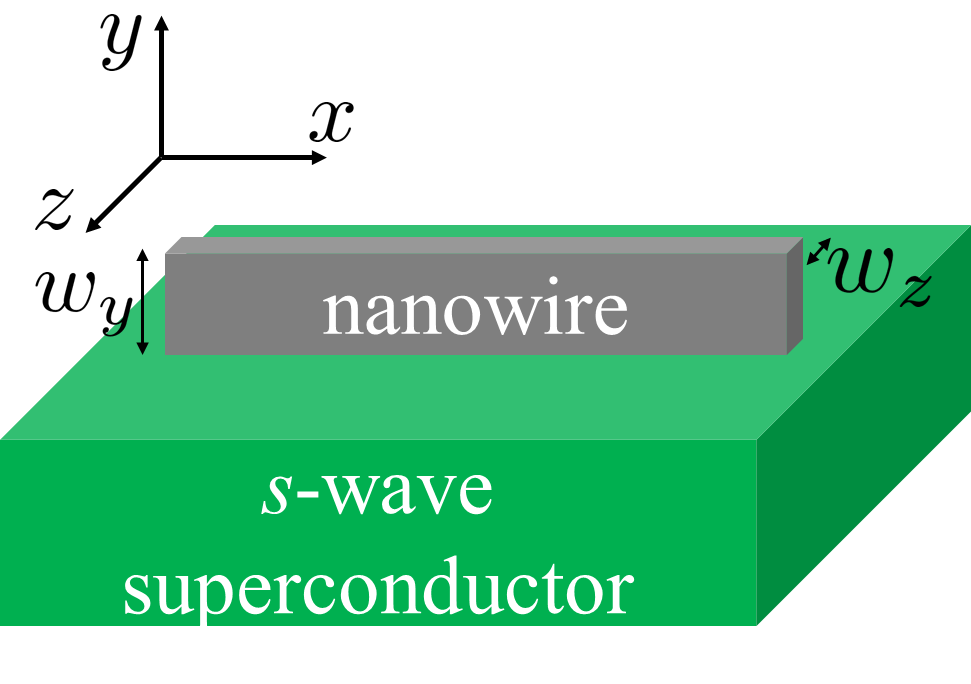}
&
\hskip -1mm
\rlap{\parbox[c]{7.5cm}{\vspace{-4.5cm}\footnotesize{(b)}}}
\includegraphics[clip=true,trim =0cm 0cm 0cm 0cm,scale=0.2]{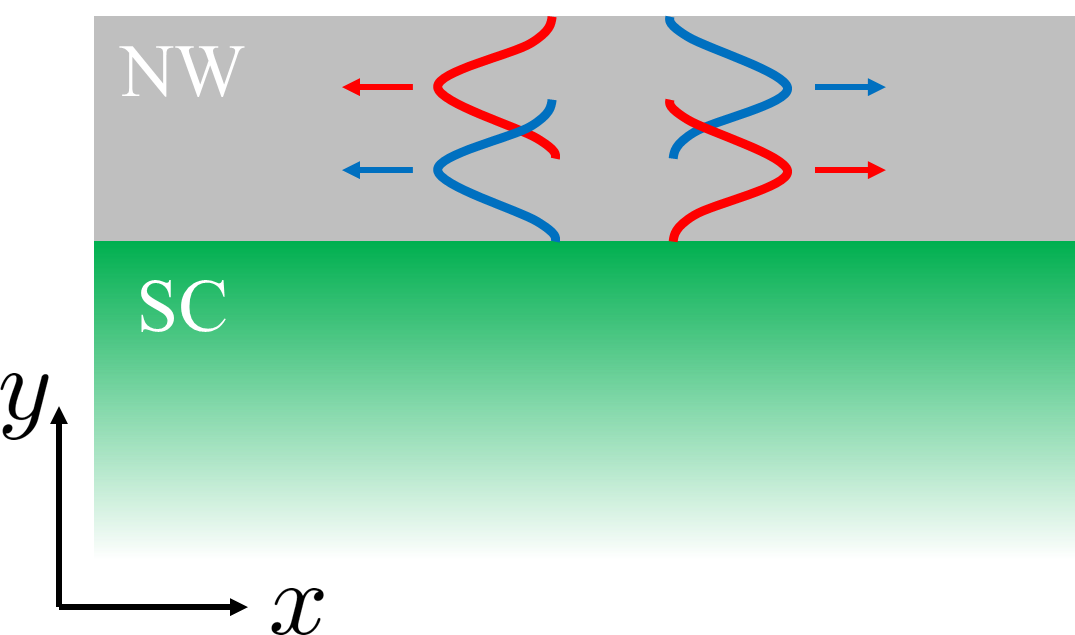}
\\
\hskip -1mm
\rlap{\parbox[c]{7.5cm}{\vspace{-4cm}\footnotesize{(c)}}}
\includegraphics[clip=true,trim =0cm 0cm 0cm -0.5cm,scale=0.38]{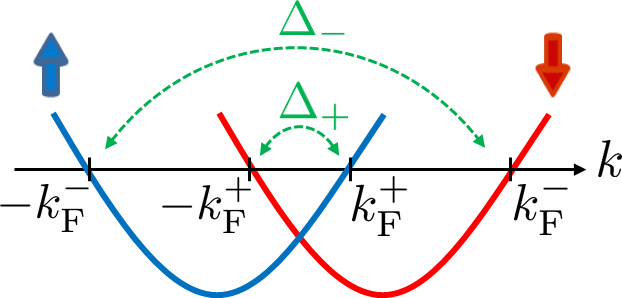}
\end{tabular}
\caption{(a) A semiconductor quasi one-dimensional nanowire on top of a bulk $s$-wave superconductor. (b) As a result of spin-orbit coupling, the spatial profile of the electronic wave functions depends on the factor $ks$, with $s=1$ for spin $\ua$, and $s=-1$ for spin $\da$, and with $k$ being the momentum in the $x$ direction. Wave functions with positive helicity ($ks>0$) are pushed towards the superconductor, while wave functions with negative helicity ($ks<0$) are pushed away from it. (c) Dispersion of the lowest transverse mode of the semiconductor nanowire. The modes near $\pm k^+_\tx{F}$ have a positive helicity and therefore experience a pairing potential $\Delta_+$ which is larger than the pairing $\Delta_-$ of modes at $\pm k^-_\tx{F}$ which have negative helicity. This results in a nonvanishing triplet pairing term $\Delta_\tx{t}=(\Delta_+-\Delta_-)/2$.}\label{fig:wire_dispersion}
\end{figure}

Consider a semiconductor spin-orbit coupled nanowire in proximity to a bulk three-dimensional $s$-wave SC as depicted in Fig.\hyperref[fig:wire_dispersion]{~\ref{fig:wire_dispersion}(a)}. The wire is infinite in the $x$ direction, while its lateral dimensions are $w_y\times w_z$. We wish to write the Hamiltonian for the lowest transverse mode of the wire. If the width of the wire is small compared to the spin-orbit coupling length, then the $z$ component of the electron's spin is approximately conserved~\cite{appendixRef}. Under this assumption, and in the absence of electron-electron interactions, the effective Hamiltonian for the lowest band is given by
\begin{equation}
\begin{split}
H_\tx{eff}=&\sum_k\left\{\sum_{ss'}\left[\left(\frac{k^2}{2m^\ast}-\mu\right)\delta_{ss'}+\alpha k\sigma^z_{ss'}\right]c^\dag_{ks}c^{\phantom{\dag}}_{ks'}\right. \\
&\hskip 3mm \left.\phantom{\sum_{ss'}}+\Delta(k)(c^\dag_{k\uparrow}c^\dag_{-k\downarrow}+{\rm h.c.})\right\},
\end{split}\label{eq:H_eff_wire}
\end{equation}
where $c^\dag_{ks}$ creates an electron in the lowest transverse mode of the wire with spin $s$ and momentum $k$ along the $x$ direction. Here $m^\ast$ is the effective mass of electrons in the wire, $\mu$ is the chemical potential, and $\alpha$ is the spin-orbit coupling strength. The induced pairing potential in the lowest transverse band is approximately given by
\begin{equation}\label{eq:Delta_k_wire}
\Delta(k) = \Delta_\tx{ind}(1+\beta k),
\end{equation}
where $\beta$ is a constant which arises due to spin-orbit interaction. Equation~\eqref{eq:Delta_k_wire} is derived in appendix~\ref{sec:low_E_derive} by perturbatively considering a general spin-orbit coupling term in the wire, and integrating out the superconductor's degrees of freedom [see Eq.~\eqref{eq:H_sm_eff_BDG}]. Physically, Eq.~\eqref{eq:Delta_k_wire} implies that modes with different helicity have a different induced pairing potential [see Fig.~\hyperref[fig:wire_dispersion]{~\ref{fig:wire_dispersion}(b-c)}]; we will elaborate on the mechanism behind this effect below.

If the chemical potential lies inside the band, then there are two pairs of Fermi points $\pm k^{+}_{\rm F}$ and $\pm k^{-}_{\rm F}$ as depicted in Fig.\hyperref[fig:wire_dispersion]{~\ref{fig:wire_dispersion}(c)}. Assuming that the induced pairing potential is much smaller than distance to the bottom of the band, we linearize the spectrum near the Fermi points as in Sec~\ref{sec:Narrow_QSHI}. This results in exactly the same Hamiltonian of Eqs.~\eqref{eq:H_lin} and~\eqref{eq:minimal_H}, where as before we define $R_{k\ua,\da} = c_{k^{\pm}_{\rm F}+k,\ua,\da}$, $L_{k\ua,\da} = c_{-k^{\mp}_{\rm F}+k,\ua,\da}$, and $\Delta_\pm = \Delta(\pm k_\tx{F}^\pm)$. The velocities of the modes at the Fermi points are given by $v_+=v_-=\sqrt{2\mu/m^\ast+\alpha^2}\equiv\bar{v}$, and the Fermi momenta are given by $k^{\pm}_\tx{F}=m^\ast(\bar{v}\mp\alpha)$. The momentum dependence of $\Delta(k)$ results in the following singlet and triplet pairing terms
\begin{equation}\label{eq:Delta_pm_wire}
  \Delta_+=\Delta_\tx{ind}(1+\beta k_\tx{F}^+) \hskip 3mm ; \hskip 3mm
  \Delta_-=\Delta_\tx{ind}(1-\beta k_\tx{F}^-),
\end{equation}
which translate into
\begin{equation}\label{eq:t_vs_s_wire}
  \Delta_\tx{s}=(1-\beta\alpha m^\ast)\Delta_\tx{ind} \hskip 3mm ; \hskip 3mm \Delta_\tx{t}= \beta m^\ast\bar{v}\Delta_\tx{ind}.
\end{equation}
Equation~\eqref{eq:Delta_k_wire} was derived in a perturbative treatment, and therefore it is valid only for sufficiently small $\beta$, for which $\Delta_\tx{s}$ exceeds $\Delta_\tx{t}$. This holds more generally, as the bare induced triplet pairing potential has to be smaller than the singlet term in the absence of interactions~\cite{Gaidamauskas2014majorana,Haim2016WithComment}.

The form of $\Delta(k)$ is derived in Appendix~\ref{sec:low_E_derive} in detail, however the essence of that derivation can be captured in the following simplified model. Let us consider the electrons in the wire to be confined in the $y$ direction by a harmonic potential $V_\tx{c}(y)=m^\ast\omega_\tx{c}^2y^2/2$, where $y=0$ is at the center of the wire. The spin-orbit coupling in the wire contributes a term of the form $\mc{H}_\tx{so}=u\partial_y V_\tx{c}(y) \hat{p}_x\sigma^z$. Ignoring the $z$ direction for the moment (justified when $w_z\ll w_y$), the electrons in the wire are governed by the first-quantized Hamiltonian
\begin{equation}\label{eq:H_wire_simplified}
\mc{H}_\tx{wire}=\frac{\hat{p}_x^2+\hat{p}_y^2}{2m^\ast} + \frac{1}{2}m^\ast\omega_c^2(y+u\hat{p}_x\sigma^z)^2.
\end{equation}
The eigenfunctions of $\mc{H}_\tx{wire}$ are $\exp(ikx)\eta_n(y+uks)$, where $s$ is the spin, and $\eta_n(y)$ are the eigenfunctions of an harmonic oscillator of mass $m^\ast$ and frequency $\omega_c$. It is now apparent that states with $ks>0$ are shifted towards the SC ($y<0$), while states with negative $ks<0$ are shifted away from the SC ($y>0$)~\cite{Moroz1999Effect}. This is illustrated in Fig.~\hyperref[fig:wire_dispersion]{\ref{fig:wire_dispersion}(b)} Upon coupling the SC to the wire, modes with $ks>0$ will therefore experience an induced pairing potential which is bigger than that of modes with $ks<0$, in accordance with Eq.~\eqref{eq:Delta_pm_wire}.

\section{Effect of repulsive interactions}
\label{sec:interactions}

After showing how the low-energy Hamiltonian of Eq.~\eqref{eq:minimal_H} with nonvanishing triplet pairing potential is obtained from two different microscopic models, we now study the effect of repulsive interactions. We will show, using both a mean-field analysis and weak-coupling RG, that short-range repulsive interactions effectively suppress the singlet term while strengthening the triplet term, thereby driving the system to the topological phase [cf. Eq.~\eqref{eq:top_inv_2}].

The full Hamiltonian is given by $H_0+H_\Delta+H_\tx{int}$, with $H_0$ and $H_\Delta$ given in Eq.~\eqref{eq:minimal_H}, and with
\begin{equation}\label{eq:H_int}
\begin{split}
H_\tx{int} = \int\tx{d}x &\left\{ g_1^\perp \left[ R^\dag_\ua(x)L^\dag_\da(x)R_\da(x)L_\ua(x) + \tx{h.c.}\right]\right.\\
&\hskip -1.5mm + g_2^+\rho_{\tx{R}\ua}(x)\rho_{\tx{L}\da}(x)+ g_2^-\rho_{\tx{R}\da}(x)\rho_{\tx{L}\ua}(x) \\
&\hskip -2mm + \left. g_2^\parallel\left[\rho_{\tx{R}\ua}(x)\rho_{\tx{L}\ua}(x) + \rho_{\tx{L}\da}(x)\rho_{\tx{R}\da}(x)\right] \right\},
\end{split}
\end{equation}
where $\rho_{\tx{R}s}(x)=R^\dag_s(x)R_s(x)$ and $\rho_{\tx{L}s}(x)=L^\dag_s(x)L_s(x)$. Here, $g_1^\perp$ is a backscattering interaction term, while $g_2^+$, $g_2^-$, and $g_2^\parallel$ are forward scattering interaction terms. In the absence of symmetry under inversion ($x\to-x$), the Fermi momenta are generally different, $k_\tx{F}^+\neq k_\tx{F}^-$ (see Fig.~\ref{fig:H_0_spectrum}). In this case $H_\tx{int}$ is the most general low-energy time-reversal symmetric Hamiltonian describing interaction between modes of opposite chirality. Interaction terms between modes of the same chirality can exist, however, they would not affect the RG flow (see Appendix~\ref{sec:RG_eqs_deriv}), nor would they contribute to our mean-field solution, and therefore we do not include them here.

\subsection{Mean-Field Analysis}
\label{sec:MF}

Before analyzing the effect of interactions using the renormalization group, it is instructive to study the mean-field solution. In this analysis we replace the low-energy interacting Hamiltonian by a quadratic Hamiltonian of the form of Eq.~\eqref{eq:minimal_H} with \emph{effective} pairing potentials $\bar\Delta_+$ and $\bar\Delta_-$. Upon determining $\bar\Delta_\pm$ by solving self-consistent equations [see Eq.~\eqref{eq:self_consist}], one can easily extract the topological invariant from this mean-field Hamiltonian.

The $g_2^\parallel$ term in Eq.~\eqref{eq:H_int} involves interaction between electrons of the same spin species. It will therefore not affect the pairing potentials $\Delta_\pm$, and its sole effect would be to change the effective chemical potential. Hence, we shall ignore it in the present mean-field treatment.

We begin by writing
\begin{equation}\label{eq:pair_term_resolved}
\begin{split}
&L_\da(x)R_\ua(x) \equiv \langle L_\da(x) R_\ua(x) \rangle + \delta_+(x), \\
&R_\da(x)L_\ua(x) \equiv \langle R_\da(x) L_\ua(x) \rangle + \delta_-(x).
\end{split}
\end{equation}
In the mean-field approximation we assume that the system has a superconducting order, and accordingly the averages of the pairing terms, $\langle L_\da(x) R_\ua(x) \rangle$ and $\langle R_\da(x) L_\ua(x) \rangle$, are large compared to their respective fluctuations, $\delta_+$ and $\delta_-$. We therefore substitute Eq.~\eqref{eq:pair_term_resolved} into Eq.~\eqref{eq:H_int} and retain terms only to first order in $\delta_\pm$. This results (up to a constant) in a mean-field Hamiltonian $H^\tx{MF}=H_0+H_\Delta^\tx{MF}$, with $H_0$ given in Eq.~\eqref{eq:minimal_H}, and with
\begin{equation}\label{eq:H_MF}
H^\tx{MF}_\Delta = \int \tx{d}x \left[\bar\Delta_+ R^\dag_\ua(x)L^\dag_\da(x) + \bar\Delta_- L^\dag_\ua(x)R^\dag_\da(x) + \tx{h.c.} \right],
\end{equation}
where
\begin{equation}\label{eq:Deltas_bar}
\begin{split}
&\bar\Delta_+=\Delta_+ + g_1^\perp\langle R_\da(x) L_\ua(x) \rangle + g_2^+\langle L_\da(x) R_\ua(x) \rangle\\
&\bar\Delta_-=\Delta_- + g_1^\perp\langle L_\da(x) R_\ua(x) \rangle + g_2^-\langle R_\da(x) L_\ua(x) \rangle.
\end{split}
\end{equation}

Since $H^\tx{MF}$ is a quadratic Hamiltonian, one can easily calculate the above pair correlation functions and arrive at self-consistent equations for $\bar\Delta_+$ and $\bar\Delta_-$. One then obtains (see Appendix~\ref{sec:self_consist})
\begin{subequations}\label{eq:self_consist}
\begin{align}
\label{eq:self_consist_a}
\begin{split}
\bar{\Delta}_+ = \Delta_+
- &\frac{g_1^\perp}{2\pi v_-} \bar{\Delta}_- \sinh^{-1}\left(v_-\Lambda/|\bar{\Delta}_-|\right)\\
- &\frac{g_2^+}{2\pi v_+} \bar{\Delta}_+ \sinh^{-1}\left(v_+\Lambda/|\bar{\Delta}_+|\right),
\end{split}\\
\label{eq:self_consist_b}
\begin{split}
\bar{\Delta}_- = \Delta_-
- &\frac{g_1^\perp}{2\pi v_+} \bar{\Delta}_+ \sinh^{-1}\left(v_+\Lambda/|\bar{\Delta}_+|\right)\\
- &\frac{g_2^-}{2\pi v_-} \bar{\Delta}_- \sinh^{-1}\left(v_-\Lambda/|\bar{\Delta}_-|\right).
\end{split}
\end{align}
\end{subequations}
These coupled equations can be solved numerically for $\bar \Delta_\pm$, after which the topological invariant of $H^\tx{MF}$ is obtained by $\mathcal{Q}=\sgn(\bar{\Delta}_+)\sgn(\bar{\Delta}_-)$. One can, however, make further analytical progress by searching for the phase boundary between $\mathcal{Q}=1$ and $\mathcal{Q}=-1$. This occurs when either $\bar{\Delta}_-=0$, or $\bar{\Delta}_+=0$. By Plugging $\bar\Delta_\pm=0$ in Eq.~\eqref{eq:self_consist}, one obtains the conditions on the parameters of the original Hamiltonian, Eqs.~\eqref{eq:minimal_H} and ~\eqref{eq:H_int}, to be on the phase boundary. If the phase boundary occurs at $\bar\Delta_+=0$, then it is described by
\begin{equation}\label{eq:phase_bound}
\frac{v_-\Lambda g_1^\perp}{|g_1^\perp \Delta_- - g_2^- \Delta_+ |} =
\sinh\left( \frac{2\pi v_-\Delta_+}{g_1^\perp \Delta_- - g_2^- \Delta_+} \right),
\end{equation}
while if it occurs at $\bar\Delta_-=0$,
\begin{equation}\label{eq:phase_bound_2}
\frac{v_+\Lambda g_1^\perp}{|g_1^\perp \Delta_+ - g_2^+ \Delta_- |} =
\sinh\left( \frac{2\pi v_+\Delta_-}{g_1^\perp \Delta_+ - g_2^+ \Delta_-} \right).
\end{equation}

\begin{figure}
    \includegraphics[clip=true,trim = 5.7mm 6mm 16.5mm 15mm,width=8.4cm]{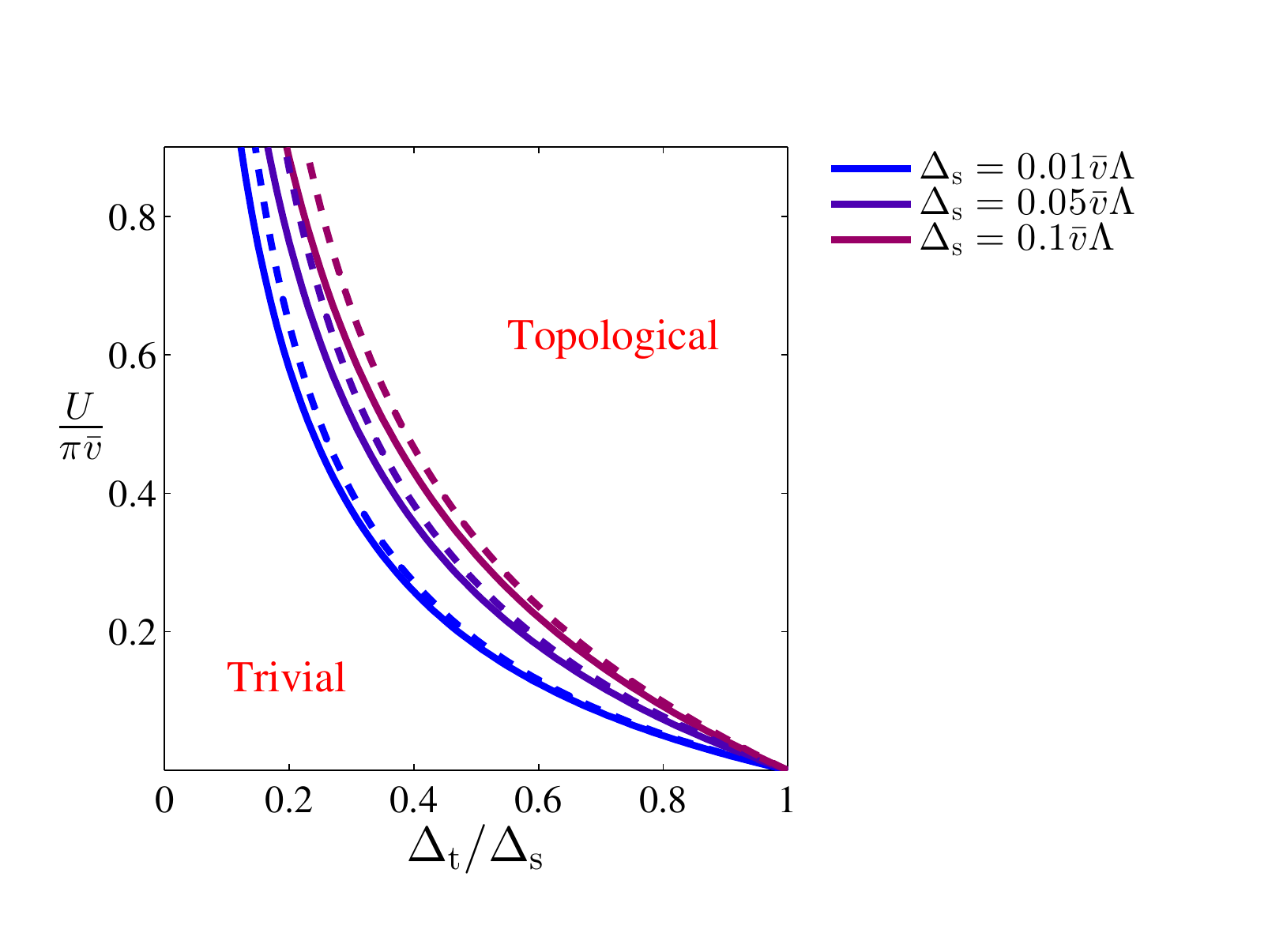}
    \caption{Phase diagram of the interacting model described in Eqs.~\eqref{eq:minimal_H} and~\eqref{eq:H_int}. The phase diagram is analyzed as a function of the interaction strength $U=g_1=g_2^+=g_2^-$, and the ratio $\Delta_\tx{t}/\Delta_\tx{s}$, for different fixed values of $\Delta_\tx{s}$. $\Delta_\tx{s}$ and $\Delta_\tx{t}$ are the singlet and triplet induced pairing potentials, respectively (referred to $\Delta^0_\tx{s,t}$ in Sec.~\ref{sec:RG}), and are related to the pairing potentials $\Delta_\pm$ through $\Delta_\tx{s,t}=(\Delta_+\pm\Delta_-)/2$. The solid lines are the phase boundaries calculated using weak-coupling RG, while the dashed lines are those calculated from Eq.~\eqref{eq:MF_ph_bound_U}, obtained from a mean-field treatment.
    Notice that for $\Delta_\tx{t}=0$ the system cannot be driven into the topological phase for any interaction strength, i.e., some initial triplet pairing term is required. For a nonzero $\Delta_\tx{t}$, the system goes through a topological phase transition at a finite value of $U$ which increases with $\Delta_\tx{s}$. For $\Delta_\tx{t}=\Delta_\tx{s}$ the system is on the verge of becoming topological, and any finite interaction will drive it to the topological phase.
    }
    \label{fig:phase_diagram_U}
\end{figure}

As a relevant example we can consider a Hubbard-type interaction, $g_1=g_2^+=g_2^-=U$, and furthermore $v_+=v_-=\bar{v}$. Let us assume without loss of generality that $|\Delta_+|>|\Delta_-|$. This means that the phase boundary will occur when $\bar{\Delta}_-=0$, namely when
\begin{equation}\label{eq:MF_ph_bound_U}
\frac{U}{\pi \bar{v}} = \frac{\Delta_\tx{s}/\Delta_\tx{t} - 1}{\sinh^{-1}\left( \bar{v}\Lambda/2|\Delta_\tx{t}| \right)}.
\end{equation}
Figure~\ref{fig:phase_diagram_U} presents the topological phase diagram, obtained using Eq.~\eqref{eq:MF_ph_bound_U} (see dashed line), as a function of $U$ and the ratio $\Delta_\tx{t}/\Delta_\tx{s}$, for different values of $\Delta_s$. As expected, for $\Delta_\tx{t}/\Delta_\tx{s}\to0$ no finite amount of interactions can bring the system to the topological phase. In contrast, when $\Delta_\tx{t}=\Delta_\tx{s}$, the system is already at a phase transition, and any nonzero $U$ suffices to drive the system to the topological phase. In the intermediate regime, the system will become topological for some finite interaction strength which increases with $\Delta_\tx{s}$.

\subsection{Weak-Coupling RG}
\label{sec:RG}

In this section we study the full interacting Hamiltonian $H_0 + H_\Delta + H_\tx{int}$, given in Eqs.~\eqref{eq:minimal_H} and \eqref{eq:H_int}, using the renormalization group (RG). We are interested in the RG flow close to the noninteracting fixed point of free electrons, described by $H_0$. Both the singlet and triplet induced pairing potentials are relevant perturbations to $H_0$, namely this is an unstable fixed point. Below we show that the introduction of $H_\tx{int}$ causes the instability to be more towards triplet pairing compared to singlet pairing.

To derive the flow equations of the various terms in $H_\Delta$ and $H_\tx{int}$ we use perturbative momentum shell Wilsonian RG for Fermions~\cite{Shankar1994Renormalization}. This procedure, whose details are given in Appendix~\ref{sec:RG_eqs_deriv}, results in
\begin{subequations}\label{eq:flow_eqs}
\begin{align}
&\dot{y}_1^\perp = -y_2 y_1^\perp ,\label{eq:KT_1}
\\
&\dot{y}_2 = -\frac{1}{2}\left(\frac{\bar{v}^2}{v_+v_-}+1\right) {y_1^\perp}^2 ,\label{eq:KT_2}
\\
&\dot{y}_2^+ = -\frac{1}{2}\frac{\bar{v}^2}{v_+v_-} {y_1^\perp}^2 ,\label{eq:y2_p_flow}
\\
&\dot{y}_2^- = -\frac{1}{2}\frac{\bar{v}^2}{v_+v_-} {y_1^\perp}^2 ,\label{eq:y2_pp_flow}
\\
&\dot{\Delta}_+ = \left( 1 - \frac{1}{2} y_2^+ \right)\Delta_+ - \frac{1}{2}\frac{\bar{v}}{v_-} y_1^\perp \Delta_- ,
\label{eq:Delta_p_flow}\\
&\dot{\Delta}_- = \left( 1 - \frac{1}{2} y_2^- \right)\Delta_- - \frac{1}{2}\frac{\bar{v}}{v_+} y_1^\perp \Delta_+ ,\label{eq:Delta_m_flow}
\end{align}
\end{subequations}
where we have defined $\bar{v}=(v_++v_-)/2$, and the dimensionless couplings $y_1^\perp=g_1^\perp/\pi\bar{v}$, $y_2^+=g_2^+/\pi v_+$, $y_2^-=g_2^-/\pi v_-$, and $y_2=g_2^+/2\pi v_+ + g_2^-/2\pi v_- - g_2^\parallel/\pi\bar{v}$. The above equations have been derived using a perturbative treatment and they are valid when $y_1$, $y_2^\parallel$, $y_2^\pm$ and $\Delta_\pm/v_\pm\Lambda$ are all smaller than 1.

Equations~(\ref{eq:KT_1},\ref{eq:KT_2}) give rise to a Kosterlitz-Thouless (KT) type of flow for $y_1^\perp$ and $y_2$. It is described by the constant of motion $A^2=y_2^2 - y_1^2$, where $y_1\equiv y_1^\perp\sqrt{(\bar{v}^2/v_+v_- + 1)/2}$. Of greatest interest for us is the region $y_2 > y_1\ge0$, this corresponds to an interaction which is repulsive on all length scales. In this case, the flow of $y_1$ and $y_2$ is given by
\begin{subequations}\label{eq:KT_flow}
\begin{align}
    y_1(\ell) &=  A \operatorname{csch} \left[ A
   \ell + \operatorname{arcoth} \frac {y_2(0)} {A}
     \right], \label{eq:y_1_of_ell}
     \\
    y_2(\ell) &= A  \coth \left[ A
    \ell + \operatorname{arcoth} \frac {y_2(0)} {A}
     \right]. \label{eq:y_2_of_ell}
\end{align}
\end{subequations}
Both $y_1$ and $y_2$ flow down, saturating after an RG time $\ell_\tx{sat}\sim A^{-1}$, at $0$ and $A$, respectively. One can insert these solutions into Eqs.~\eqref{eq:y2_p_flow} and \eqref{eq:y2_pp_flow}, and integrate to obtain $y_2^+$ and $y_2^-$, respectively. The interaction couplings $y_1^\perp$, $y_2^+$, and $y_2^-$ can then be inserted into Eqs.~(\ref{eq:Delta_m_flow},\ref{eq:Delta_p_flow}) which generally require a numerical solution for $\Delta_\pm$.

We wish to determine the topological phase diagram of the system as a function of its initial couplings. We solve the above flow equations up to an RG time $\ell^\ast$, at which one of the pairing potential flows to strong coupling, namely $|\Delta_\pm(\ell^\ast)|/v_\pm\Lambda=1$. Beyond this point the perturbative RG treatment is not valid anymore. Let us assume, without loss of generality, that $\Delta_+$ flows to strong coupling first. This in particular means that the interaction couplings (which have flown down) are small in comparison to it, namely  $y_1^\perp, y_2^\parallel,y_2^\pm \ll |\Delta_+(\ell^\ast)|/v_+\Lambda=1$. If at this point $\Delta_-(\ell^\ast)/v_-\Lambda$ happens also to be large in comparison to $y_1^\perp, y_2^\parallel,y_2^\pm$, then we can neglect the interaction couplings. One can then use the topological invariant of a noninteracting system [see Eq.~\eqref{eq:top_inv}], $\mc{Q} = \sgn[\Delta_+(\ell^\ast)]\sgn[\Delta_-(\ell^\ast)]$. Generally, however, $\Delta_-(\ell^\ast)$ can be small, and one has to modify the expression for $\mc{Q}$ to account for the non-negligible interaction terms.

To this end we note that since $\Delta_+(\ell^\ast)$ is large, the positive-helicity degrees of freedom [$R_\ua(x)$ and $L_\da(x)$] are gapped, and we can safely integrate them out. Upon doing that, one is left with an action containing only the negative-helicity fields [$R_\da(x)$ and $L_\ua(x)$], with a pairing potential $\Delta'_-=\Delta_-(\ell^\ast)+\delta\Delta_-$. To leading order in the interaction couplings, the correction is given by (see appendix~\ref{sec:RG_eqs_deriv})
\begin{equation}\label{eq:correc_to_Delta_minus}
\begin{split}
\delta\Delta_- &= -\frac{\bar{v}}{2v_+}y_1^\perp(\ell^\ast)\Delta_+(\ell^\ast)\sinh^{-1}\left[\frac{v_+\Lambda}{|\Delta_+(\ell^\ast)|}\right]=\\
&= -\frac{1}{2}y_1^\perp(\ell^\ast)\sgn[\Delta_+(\ell^\ast)]\sinh^{-1}(1)\bar{v}\Lambda.
\end{split}
\end{equation}
At this point we can continue the RG procedure, applied only to the negative-helicity degrees of freedom,
\begin{subequations}
\begin{align}\label{eq:flow_eqs_ssecnd_stp}
\dot{y}_2^- &= 0,\\
\dot\Delta'_- &= \left( 1 - \frac{1}{2} y_2^- \right)\Delta'_-,
\end{align}
\end{subequations}
namely $\Delta'_-$ flows to strong coupling (without changing sign), while $y_2^-$ remains perturbative. We can therefore use the topological invariant of noninteracting systems, only with $\Delta_-(\ell^\ast)$ substituted by $\Delta'_-$, $\mc{Q}=\sgn[\Delta_+(\ell^\ast)]\sgn[\Delta'_-]$. Finally, accounting also for the possibility that $\Delta_-$ flows to strong coupling before $\Delta_+$, we can write
\begin{equation}\label{eq:top_inv_RG}
\begin{split}
\mc{Q} = &\sgn\left\{\frac{\Delta_+(\ell^\ast)}{\bar{v}\Lambda} - \frac{\sinh^{-1}(1)}{2}y_1^\perp(\ell^\ast)\sgn[\Delta_-(\ell^\ast)]\right\}\times \\
&\sgn\left\{\frac{\Delta_-(\ell^\ast)}{\bar{v}\Lambda} - \frac{\sinh^{-1}(1)}{2}y_1^\perp(\ell^\ast)\sgn[\Delta_+(\ell^\ast)]\right\},
\end{split}
\end{equation}
where $\ell^\ast$ is the RG time when the first of $\Delta_+$ and $\Delta_-$ reaches strong coupling.

To understand how repulsive interactions drive the system into the TRITOPS phase, let us concentrate on the special case, $v_+=v_-$, $y_2^-=y_2^+$, for which Eqs.~(\ref{eq:Delta_m_flow},\ref{eq:Delta_p_flow}) reduce to
\begin{subequations}\label{eq:Delta_s_t_flow_eqs}
\begin{align}
&\dot{\Delta}_\tx{s} = \left( 1 - \frac{1}{2} y_2^+ - \frac{1}{2} y_1 \right)\Delta_\tx{s},
\label{eq:Delta_s_flow}\\
&\dot{\Delta}_\tx{t} = \left( 1 - \frac{1}{2} y_2^+ + \frac{1}{2} y_1 \right)\Delta_\tx{t}.
\label{eq:Delta_t_flow}
\end{align}
\end{subequations}
The effect of forward scattering and of backscattering on the pairing potentials is now apparent. The forward scattering term $y_2^+$ equally suppresses the singlet and triplet pairing terms. The backscattering term $y_1$, on the other hand, suppresses $\Delta_\tx{s}$, while strengthening $\Delta_\tx{t}$, causing the latter to flow faster to strong coupling. From Eq.~\eqref{eq:Delta_s_t_flow_eqs} one can extract the ratio between the triplet and singlet pairing terms as a function of RG time,
\begin{equation}\label{eq:r_t_s_ell}
    \frac {\Delta_{\mathrm{t}}(\ell)} {\Delta_{\mathrm{s}}(\ell)} = \frac{\Delta_{\mathrm{t}}^0} {\Delta_{\mathrm{s}}^0}
    \exp\left[\int_{0}^{\ell}\!\mathrm{d}\ell'y_1(\ell')\right].
\end{equation}

If the time it takes $y_1$ to flow to zero, $\ell_\tx{sat}$, is much shorter than $\ell^\ast$, we can approximate the ratio $\Delta_\tx{t}(\ell^\ast)/\Delta_\tx{s}(\ell^\ast)$ by taking the upper limit of the above integral to infinity. Using Eq.~\eqref{eq:y_1_of_ell}, one obtains in this case
\begin{equation}\label{eq:Long_RG_time_approx}
    \frac {\Delta_\mathrm{t}(\ell^\ast)} {\Delta_{\mathrm{s}}(\ell^\ast)} \simeq\frac
    {\Delta_{\mathrm{t}}^0} {\Delta_{\mathrm{s}}^0}\sqrt{\frac {y_2^0+y_1^0}{y_2^0-y_1^0} }.
\end{equation}
Furthermore, since by our assumption $y_1(\ell^\ast)\simeq 0$ (follows from $\ell_\tx{sat}\ll\ell^\ast$), Eq.~\eqref{eq:top_inv_RG} tells us that the condition for the system to be topological is simply $|\Delta_\tx{t}(\ell^\ast)|>|\Delta_\tx{s}(\ell^\ast)|$. We wish to understand when this approximation is valid. To this end, we can estimate the time it would take for one of the pairing potentials to reach strong coupling, $\ell^\ast\sim \ln(v_\pm\Lambda/\Delta_\pm^0)$~\cite{EstimateEllStar}. Namely, the above long RG-time approximation will be valid if the initial pairing potentials are small enough such that $\Delta_\pm^0 \ll v_\pm\Lambda \exp(-1/A)$. Note that the above approximation will necessarily be violated close to the separatrix of the KT flow, since there $A\to 0$.

\begin{figure}[t]
    \hskip -5mm
    \includegraphics[clip=true,trim = 45mm 0mm 50mm 5mm,width=0.4\textwidth]{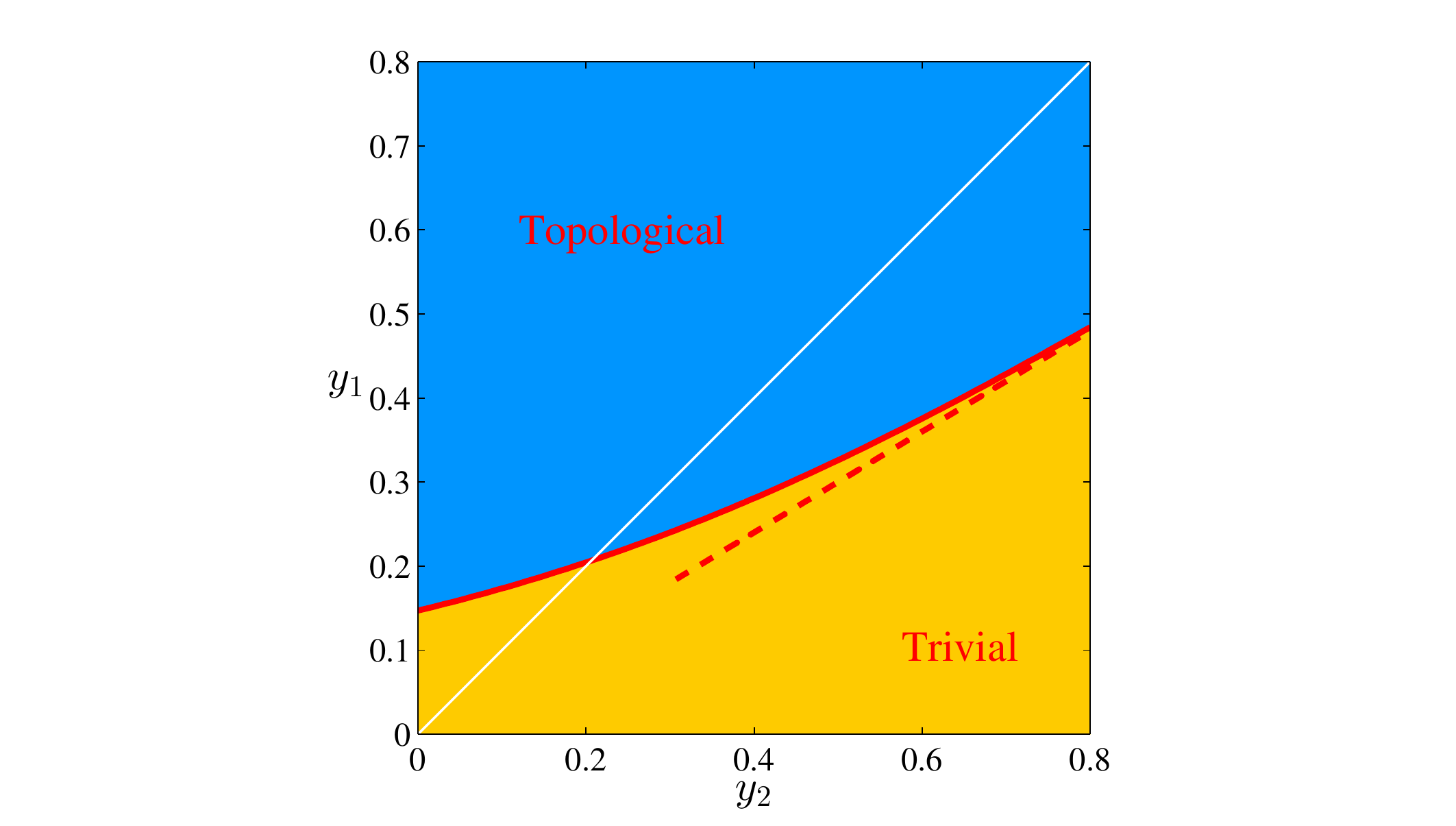}
    \caption{Phase diagram of the interacting model described in Eqs.~\eqref{eq:minimal_H} and~\eqref{eq:H_int}, as a function of forward ($y_2$) and backward ($y_1$) scattering interaction terms. The solid red line shows the phase boundary, calculated using the RG flow equations, Eq.~\eqref{eq:flow_eqs}, and the topological invariant, Eq.~\eqref{eq:top_inv_RG}. The dashed red line indicates the long RG-time approximation for the phase boundary, Eq.~\eqref{eq:phaseBoundaryEquation}. It agrees with the numerical result when $\Delta_\pm^0 \ll v_\pm\Lambda \exp(-1/A)$, where $A^2=y_2^2-y_1^2$. The white solid line corresponds to the separatrix of the Kosterlitz- Thouless flow, above which $y_1$ and $y_2$ flow to strong coupling. In obtaining this phase diagram, we have used $v_-=v_+=\bar{v}$, $y_2^+=y_2^-=y_2$, and the initial singlet and triplet pairing potentials were taken to be $\Delta^0_{\mathrm{t}}=0.01\bar{v}\Lambda$ and $\Delta^0_{\mathrm{s}}=0.02\bar{v}\Lambda$, respectively.}
        \label{fig:phasediagram}
\end{figure}

We can now use the result, Eq.~\eqref{eq:Long_RG_time_approx} to construct the
phase diagram of the system as a function of the initial values of $y_2$ and $y_1$, given fixed
initial conditions for $\Delta_{\mathrm{s}}$ and $\Delta_\mathrm{t}$. Assuming that we can take the long RG-time limit, we can find an equation for the phase boundary in the  $y_2y_1-$plane, by setting equation \eqref{eq:Long_RG_time_approx} to $1$ and solving for $y_1$. One then immediately finds that the phase boundary obeys the equation
\begin{equation}\label{eq:phaseBoundaryEquation}
y_1^0 = \frac {1 - (\Delta^0_\tx{t}/\Delta^0_\tx{s})^2}{1 + (\Delta^0_\tx{t}/\Delta^0_\tx{s})^2}\cdot y_2^0,
\end{equation}
namely, the system is in the topological phase above this line in the $y_1y_2-$plane. The topological region becomes bigger as the ratio $\Delta_\tx{t}^0/\Delta_\tx{s}^0$ increases. In Fig.~\ref{fig:phasediagram} we present the topological phase diagram in the $y_2y_1$-plane for fixed initial values $\Delta_\tx{s}$ and $\Delta_\tx{t}$. The phase boundary is obtained by numerically solving Eq.~\eqref{eq:flow_eqs} up to a time $\ell^\ast$, and then invoking Eq.~\eqref{eq:top_inv_RG}, with $\ell^\ast$ being the RG time when the first coupling becomes 1. The dashed red line shows the long-RG time approximation of the phase boundary, Eq.~\eqref{eq:phaseBoundaryEquation}. As anticipated, it becomes more accurate as $A$ increases. We note that above the separatrix of the KT flow, $y_1$ and $y_2$ flow to strong coupling and the system is driven into an intrinsically topological phase~\cite{Keselman2015gapless,kainaris2016interaction}, irrespective of the initial induced potentials $\Delta_\pm$. Some nonvanishing induced pairing is however necessary to keep the system fully gapped.

Let us now reconsider the case of a Hubbard-type interaction, $g_1^\perp=g_2^+=g_2^-=U$, and $g_2^\parallel=0$. Note that for $v_+=v_-$ this mean $y_2=y_1$, while for $v_+\neq v_-$, this means $y_2\ge y_1$ [see the definitions below Eq.~\eqref{eq:flow_eqs}]. Importantly, in both cases the KT flow equations dictates that the interaction couplings flow down. Figure~\ref{fig:phase_diagram_U} shows the phase diagram for this Hubbard-type interaction, for $v_+=v_-$. The critical interaction strength $U$ which defines the phase boundary is numerically calculated as a function of the initial ratio $\Delta^0_\tx{t}/\Delta^0_\tx{s}$, for different fixed values of $\Delta^0_\tx{s}$. We note that this phase boundary (solid lines) agrees well with that obtained from the mean field analysis (dashed lines), given in Eq.~\eqref{eq:MF_ph_bound_U}.

The results presented in Figs.~\ref{fig:phase_diagram_U} and~\ref{fig:phasediagram} are both for the case of $g_2^+=g_2^-$, $v_+=v_-$. Under these conditions, if the initial triplet term is zero, it will remain zero for all RG times, as can be seen from Eq.~\eqref{eq:Delta_t_flow}. This is no longer the case upon relaxing one of these conditions, since the flow equations generally couple $\Delta_\tx{t}$ to $\Delta_\tx{s}$ [see Eq.~\eqref{eq:flow_eqs}]. Consequently, a phase transition into the topological phase can occur at a finite interaction strength, even for vanishingly small initial $\Delta_\tx{t}^0$. Nevertheless, systems in which the initial triplet term can be of the order of the singlet term, such as those presented in Sec.~\ref{sec:Realizations}, are more susceptible to being driven into the topological phase by the effect of repulsive interactions.

\section{Discussion}
\label{sec:discussion}

We have presented and studied a general low-energy model for a one-dimensional system where the interplay between externally-induced superconductivity and repulsive Coulomb interactions stabilizes a time-reversal invariant topological superconducting phase. This phase is characterized by a Kramers' pair of zero-energy Majorana bound state at each end of the system.

We have suggested two experimentally-accessible setups of proximity-coupled systems which realize this low-energy model, and which can therefore serve as a platforms for realizing time-reversal invariant topological superconductivity. These are (i) a narrow strip of a 2d topological insulator, partially covered by an $s$-wave superconductor, and (ii) a quasi 1d semiconductor nanowire in proximity to an $s$-wave superconductor.

We expect the excitation gap of the system to protect the topological phase against a moderate amount of disorder, namely disorder with associated mean free time which is large in comparison with the inverse energy gap. This is the case for the class-D TSC~\cite{Motrunich2001Griffiths,Brouwer2011Probability}, which can be thought of as ``half'' of a class-DIII TSC (or TRITOPS).

An experimental signature of this phase can be obtained by probing the Kramers' pair of Majorana bound states which reside at each end of the system. By coupling the end of the system to a normal-metal lead, the differential conductance can be measured. At zero temperature this should yield a zero-bias peak which is quantized to $4e^2/h$~\cite{Wong2012majorana,Haim2014time,Dumitrescu2014magnetic}. The behavior of this conductance peak upon breaking time-reversal symmetry by a Zeeman field has features which are distinctive of a Majorana Kramers' pair~\cite{Keselman2013inducing,Haim2014time,Dumitrescu2014magnetic}. Alternatively, current correlations in a two-lead setup can be used to detect signatures which are unique to Majorana bound states~\cite{Haim2015signatures,Haim2015current,li2015detection}. Coulomb-blockade spectroscopy, recently applied to TSC with broken TRS, can be used to probe also the TRITOPS phase, where the topological transition is expected to be manifested in the disappearance of the even-odd effect. Experimental signatures have also been suggested to exist in the anomalous behavior of Josephson junctions involving TRITOPS~\cite{chung2013time,Zhang2014anomalous,Kane2015the}.

It is interesting to examine the strength of electron-electron interactions in the suggested experimental setups of Sec.~\ref{sec:Realizations}. Given an estimate for the induced pairing potentials, $\Delta_\tx{s}$ and $\Delta_\tx{t}$, one can then try and place a given system on the phase diagram of Fig.~\ref{fig:phase_diagram_U} to predict whether it is in the topological or trivial phase.

First we note that the Coulomb interaction between the electrons is screened by the presence of the SC. This sets a finite range for the interaction, given roughly by the lateral distance between the SC and the electrons in the system. This can be estimated as the width $d$ of the QSHI strip (or of the wire in the case of the setup in Sec.~\ref{sec:Rashba_wire}). At short electron-electron distances ($|x-x'|\ll d$) the divergence of the Coulomb interaction is regularized by the finite width of the system, $V(x-x')\sim e^2/4\pi\eps d$, where $\eps$ is the permittivity. If the Fermi wavelength is sufficiently larger than the interaction range $d$, then the forward and backward scattering interactions are of the same order,
\begin{equation}\label{eq:int_strength}
g_1^\perp,g_2^+,g_2^- \sim  d\cdot \frac{e^2}{4\pi\eps d}=\frac{e^2}{4\pi\eps},
\end{equation}
and accordingly the dimensionless interaction strength is $U/\pi\hbar\bar{v}\sim e^2/4\pi^2\hbar \bar{v}\eps$. The velocity $\bar{v}$ depends on details such as the chemical potential. However, a reasonable estimate is $\bar{v}\sim 10^5\tx{m}/\tx{s}$. Takeing $\eps\sim10\eps_0$ results in $U/\pi\hbar\bar{v}\sim 0.7$.

Based on recent experiments~\cite{mourik2012signatures,Das2012zero} one can estimate for the induced pairing potential, $\Delta_{s}\sim 0.1\tx{meV}$. The energy cutoff for the low-energy theory should be roughly given by the distance to the bottom of the band [see Figs.~\hyperref[fig:Narrow_QSHI]{\ref{fig:Narrow_QSHI}(c)} and ~\hyperref[fig:Narrow_QSHI]{\ref{fig:wire_dispersion}(c)}] which again depends on the chemical potential. Looking at the phase diagram of Fig.~\ref{fig:phase_diagram_U}, and assuming $\Delta_\tx{s}/\hbar \bar{v}\Lambda \sim 0.1$, we see that the system is expected to be in the topological phase for initial ratios $|\Delta_\tx{t}|/|\Delta_\tx{s}|$ greater than about 0.3.

\section*{ACKNOWLEDGEMENTS}
We have benefited from discussions with Y. Baum, A. Keselman, K. Michaeli, M.-T. Rieder, and Y. Schattner. E. B. was supported by the Minerva foundation,  by a Marie Curie Career Integration Grant (CIG), and by the European Research Council (ERC) under the European Union's Horizon 2020 research and innovation programme (grant agreement No. 639172). Y. O. was supported by the Israeli Science Foundation (ISF), by the Minerva foundation, by the Binational Science Foundation (BSF) and by the ERC, grant No. 340210 (FP7/2007-2013). K. F. was supported by the Danish National Research Foundation and by the Danish Council for Independent Research $|$ Natural Sciences.

\appendix

\section{Topological criterion}
\label{sec:top_criter}

Formulas for the topological invariant of 1d Hamiltonians in class DIII were derived in several previous studies~\cite{Qi2010topological,Fulga2011scattering,budich2013topological,Gaidamauskas2014majorana,Haim2014time,Mandal2015counting,Haim2016WithComment}. We shall focus here on the low-energy model described by the Hamiltonian of Eq.~\eqref{eq:minimal_H}. Namely, we are interested in the condition on the parameters of Eq.~\eqref{eq:minimal_H} for which the system is in the TRITOPS phase with a Kramers pair of Majorana bound states at each end of the system. We shall use a scattering-matrix formalism to obtain a condition for the existence of a zero energy bound state~\cite{Fulga2011scattering,Rieder2013Reentrant}.

\begin{figure}
\includegraphics[clip=true,trim =0mm 0mm 0mm 0mm,width=0.47\textwidth]{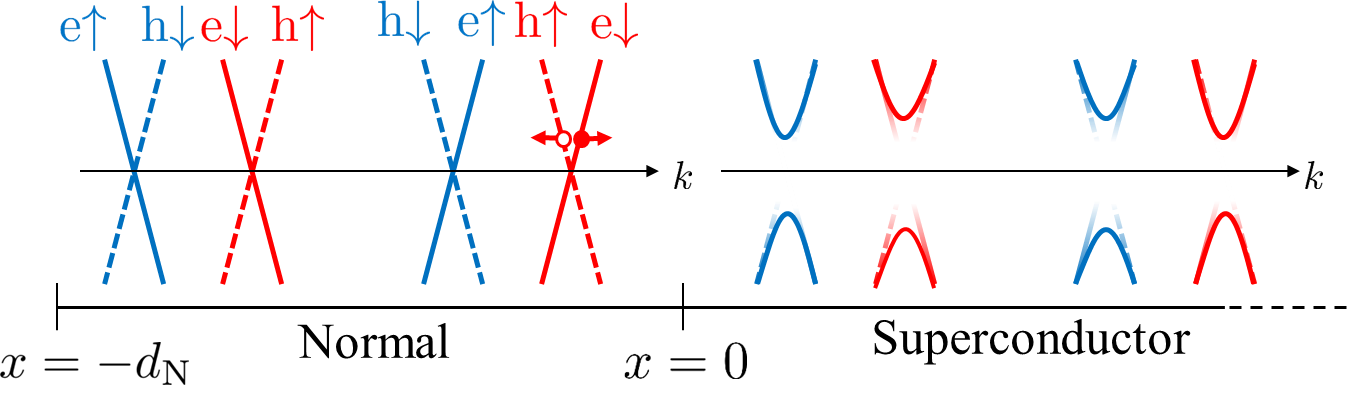}
\caption{Theoretical construction for obtaining the criterion for the low-energy Hamiltonian $H=H_0+H_\Delta$ to be in the topologically nontrivial phase [cf. Eq.~\eqref{eq:minimal_H}]. The semi-infinite region $x>0$ is described by the Hamiltonian $H=H_0+H_\Delta$, while the region $-d_\tx{N}<x\le0$ is described by $H_0$. Using the scattering matrices at $x=0$ and at $x=-d_\tx{N}$ we obtain the condition for the existence of a zero-energy bound state (In fact two bound states due to Kramers' theorem), signifying that the system is in the topological phase.}\label{fig:Top_Crit}
\end{figure}

Let our system, which is described by $H=H_0+H_\Delta$, extend from $x=0$ to $x\to \infty$. We attach on the left a normal-metal stub, extending from $x=-d_N$ to $x=0$, and described by $H_0$. This is depicted in Fig.~\ref{fig:Top_Crit}. In the absence of a barrier at $x=0$, a spin-$\ua$ ($\da$) electron incident from the left at subgap energies is Andreev reflected as a hole with spin $\da$ ($\ua$), with an amplitude $a_+$ ($a_-$), where~\cite{andreev1966electron,Beenakker1991universal}
\begin{equation}
a_\pm(E) = \frac{E-i\sqrt{\Delta_\pm^2-E^2}}{\pm\Delta_\pm},
\end{equation}
for $E\le\Delta_\pm$, as can be checked by matching the wave functions at $x=0$. The reflection matrix at the $x=0$ interface is then given by
\begin{equation}\label{eq:r_NS}
r_\tx{NS} = \begin{pmatrix}0&A^\ast(-E)\\A(E)&0\end{pmatrix} \hskip 1mm ; \hskip 1mm
A= \begin{pmatrix}0&a_-(E)\\a_+(E)&0\end{pmatrix}.
\end{equation}
At the end of the stub, $x=-d_N$, electrons and holes experience total normal reflection. The reflection matrix can therefore be written most generally as
\begin{equation}
r_\tx{N} = \begin{pmatrix}R(E)&0\\0&R^\ast(-E)\end{pmatrix} \hskip 1mm ; \hskip 1mm
R= \begin{pmatrix}e^{i\alpha(E)}&0\\0&e^{i\alpha(E)}\end{pmatrix},
\end{equation}
where $\alpha(E)$ is a phase which includes also the phase acquired during the propagation in the metallic region. The form of $r_N$ is dictated by particle-hole symmetry, while the form of $R(E)$ is dictated by time-reversal symmetry, $R(E)=\sigma^yR^\tx{T}(E)\sigma^y$, and by its unitarity.

Upon being reflected, once at $x=0$ and once at $x=-d_\tx{N}$, the wave function must comeback to itself. This implies a condition for the existence of a bound state
\begin{equation}
\det(1-r_\tx{N} r_\tx{NS})=0.
\end{equation}
At zero energy this reduces to
\begin{equation}\label{eq:det_cond}
|1-a^\ast_+a_-|^2=0,
\end{equation}
and finally, since at zero energy $a_\pm(0)=-i\sgn(\Delta_\pm)$, the condition for having a zero-energy bound state is
\begin{equation}
\sgn(\Delta_+)\sgn(\Delta_-)=-1.
\end{equation}
Notice that the power of $2$ in Eq.~\eqref{eq:det_cond} signifies that there are indeed \emph{two} zero-energy solutions, these are the Kramers' pair of Majorana bound states.

\section{Self-consistent equations}
\label{sec:self_consist}

We derive here the self-consistent mean-field equations~\eqref{eq:self_consist_a} and~\eqref{eq:self_consist_b}, by calculating the correlation functions in Eq.~\eqref{eq:Deltas_bar}. To this end we write the mean-field Hamiltonian, defined above Eq.~\eqref{eq:H_MF}, in momentum space
\begin{equation}\label{H_MF_k}
\begin{split}
H^\tx{MF} = \sum_{|k|<\Lambda}&\left\{(R_{k\ua}^\dag ,L_{-k\da})\begin{pmatrix}v_+k&\bar\Delta_+\\ \bar\Delta_+&-v_+k\end{pmatrix}\begin{pmatrix}R_{k\ua}\\L_{-k\da}^\dag\end{pmatrix}\right.+\\
&\left.+(L_{-k\ua}^\dag ,R_{k\da})\begin{pmatrix}v_-k&\bar\Delta_-\\ \bar\Delta_-&-v_-k\end{pmatrix}\begin{pmatrix} L_{-k\ua} \\ R_{k\da}^\dag \end{pmatrix}\right\},
\end{split}
\end{equation}
where $R_s(x) = (1/\sqrt{l})\sum_{|k|<\Lambda}R_{ks}\exp(-ikx)$ and $L_s(x) = (1/\sqrt{l})\sum_{|k|<\Lambda}L_{ks}\exp(-ikx)$, $l$ being the length of the system, and $\Lambda$ being the high momentum cutoff of the theory. $H^\tx{MF}$ can be readily diagonalized, yielding
\begin{equation}
H^\tx{MF}=E_\tx{G} + \sum_{|k|<\Lambda}\sum_{\tau=\pm} E_{k\tau}(\alpha^\dag_{k\tau}\alpha_{k\tau}+ \beta^\dag_{k\tau}\beta_{k\tau}),
\end{equation}
with $E_{k\pm}=\sqrt{\bar\Delta_\pm^2 + (v_\pm k)^2}$, and with $\alpha_{k\pm}$ and $\beta_{k\pm}$ given by
\begin{subequations}\label{eq:Ferm_trans}
\begin{align}
\begin{pmatrix}\alpha_{k+}\\ \beta^\dag_{k+}\end{pmatrix}=
\begin{pmatrix} \cos\theta_{k+} & \sin\theta_{k+} \\ \sin\theta_{k+} & -\cos\theta_{k+} \end{pmatrix}
\begin{pmatrix}R_{k\ua} \\ L^\dag_{-k\da} \end{pmatrix}, \\
\begin{pmatrix}\alpha_{k-}\\ \beta^\dag_{k-}\end{pmatrix}=
\begin{pmatrix} \cos\theta_{k-} & \sin\theta_{k-} \\ \sin\theta_{k-} & -\cos\theta_{k-} \end{pmatrix}
\begin{pmatrix}L_{-k\ua} \\ R^\dag_{k\da} \end{pmatrix} ,
\end{align}
\end{subequations}
where
$\cos(2\theta_{k\pm})=v_\pm k/\sqrt{\bar\Delta_\pm^2+(v_\pm k)^2}$ and $\sin(2\theta_{k\pm})=\bar\Delta_\pm/\sqrt{\bar\Delta_\pm^2+(v_\pm k)^2}$.

By inverting Eq.~\eqref{eq:Ferm_trans}, and using the fact that $\alpha_{k\pm}$ and $\beta_{k\pm}$ annihilate the ground state of $H^\tx{MF}$, one obtains (at zero temperature)
\begin{equation}\label{eq:pairing_corr_plus}
\begin{split}
&\langle L_\da(x) R_\ua(x) \rangle = \frac{1}{l}\sum_{|k|<\Lambda} \langle L_{-k\da} R_{k\ua} \rangle = \\
&= -\frac{1}{2l}\sum_{|k|<\Lambda} \sin(2\theta_{k+}) = - \frac{\bar\Delta_+}{4\pi}\int_{-\Lambda}^\Lambda \frac{\tx{d}k}{\sqrt{\bar\Delta_+^2+(v_+ k)^2}}= \\
&= -\frac{\bar\Delta_+}{2\pi v_+}\sinh^{-1}\left({v_+\Lambda/|\bar\Delta_+|}\right),
\end{split}
\end{equation}
and similarly
\begin{equation}\label{eq:pairing_corr_minus}
\langle R_\da(x) L_\ua(x) \rangle = -\frac{\bar\Delta_-}{2\pi v_-}\sinh^{-1}\left({v_-\Lambda/|\bar\Delta_-|}\right).
\end{equation}
Inserting Eqs.~\eqref{eq:pairing_corr_plus} and~\eqref{eq:pairing_corr_minus} in Eq.~\eqref{eq:Deltas_bar} results in the self-consistent equations for $\bar\Delta_\pm$, Eqs.~\eqref{eq:self_consist_a} and~\eqref{eq:self_consist_b}.

\section{Derivation of the low-energy wire Hamiltonian}
\label{sec:low_E_derive}

In this appendix we derive the low-energy Hamiltonian for a Rashba spin-orbit coupled wire in proximity to a three-dimensional $s$-wave SC. We show that it has the form of Eq.~\eqref{eq:H_eff_wire} with a momentum-dependent pairing potential $\Delta(k)$. This results in a nonvanishing triplet pairing term which, as explained in the main text, makes the system susceptible to being driven into a topological phase in the presence of strong enough repulsive interactions.

We consider an infinite quasi one-dimensional wire with lateral dimensions $w_y\gg w_z$. As depicted in Fig.~\hyperref[fig:wire_dispersion]{\ref{fig:wire_dispersion}(a)}, the wire is placed on the surface of a conventional $s$-wave SC along the $x$ axis in the plane defined by $y=-w_y/2$. The Hamiltonian for the wire in first quantization is given by
\begin{equation}
\mathcal{H}_{\rm sm} = -\frac{\hat{\nabla}^2}{2m_{\rm sm}}-i\boldsymbol\lambda(y,z)\cdot(\boldsymbol{\sigma}\times\nabla)+V_{\rm c}(y,z)
\label{eq:H_sm}
\end{equation}
where $m_{\rm sm}$ is the effective mass of electrons in the semiconductor wire, $V_{\rm c}(y,z)$ is the confining potential to be described below, and $\boldsymbol\lambda(y,z)$ is a spin-orbit coupling field which stems from the internal effective electric field felt by the conduction electrons in the wire. Here, $\boldsymbol{\sigma}$ is a vector of Pauli matrices in spin space. The SC is described by the Hamiltonian
\begin{equation}
\begin{split}
H_{\rm sc}&=H_{\rm N}+H_{\Delta},\\
H_{\rm N}&=\sum_{s=\uparrow,\downarrow}\int {\rm d}^3\boldsymbol{r}\,\psi_s^\dag(\boldsymbol{r})\left[\frac{-\nabla^2}{2m_\tx{sc}}-\mu_\tx{sc}\right]\psi_s(\boldsymbol{r}),\\
H_\Delta&=\int {\rm d}^3\boldsymbol{r}\,\Delta_{\rm sc}\psi_{\uparrow}^\dag(\boldsymbol{r})\psi_{\downarrow}^\dag(\boldsymbol{r})+{\rm h.c.},
\end{split}\label{eq:SC_Hamiltonian}
\end{equation}
where $\mu_{\rm sc}$ is the chemical potential, $m_{\rm sc}$ is the effective mass of electrons in the normal state of the SC, $\Delta_{\rm sc}$ is the superconducting gap, and $\psi_s^\dag$ is a creation operator of electrons with spin $s=\uparrow,\downarrow$ in the SC.

\begin{figure}
\includegraphics[clip=true,trim =0cm 0cm 0cm 0cm,width=0.37\textwidth]{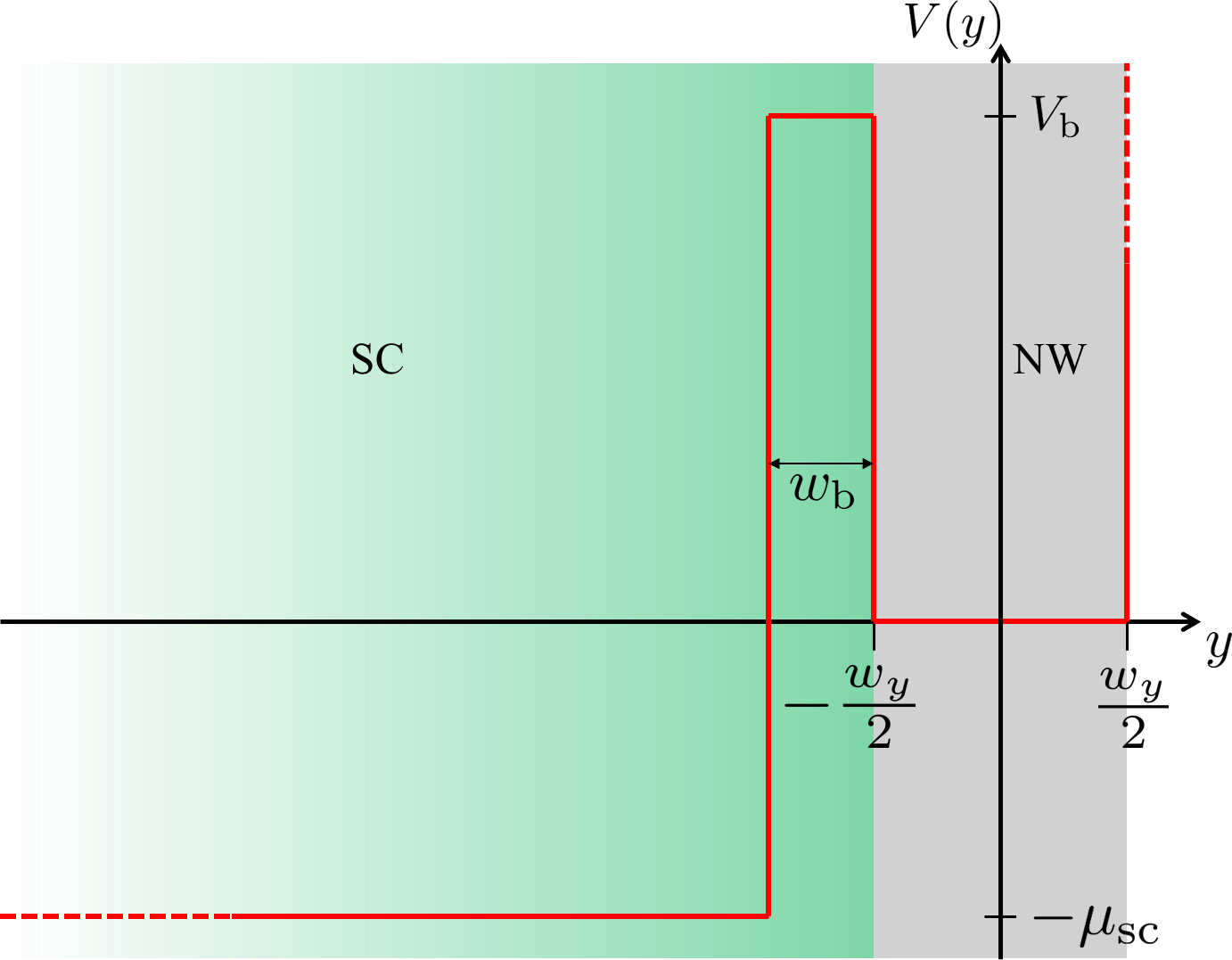}
\caption{The profile of the electronic confining potential (red line) projected along the $y$ direction for the system depicted in Fig.~\hyperref[fig:wire_dispersion]{\ref{fig:wire_dispersion}(a)}. In order to construct a tunneling Hamiltonian we consider a thin insulating layer between the nanowire and the superconductor. This is described by a potential barrier of height $V_\tx{b}$ and width $w_\tx{b}$.}\label{fig:NW_and_conf_pot}
\end{figure}

Our goal is to derive an effective low-energy Hamiltonian for the semiconductor nanowire. To this end we first construct a tunneling Hamiltonian by following Bardeen's transfer Hamiltonian approach~\cite{Bardeen1961tunneling}, and then integrate out the superconductor's degrees of freedom. As we show below, the spin-orbit coupling term in Eq.~\eqref{eq:H_sm} modifies the form of the induced pair potential in the wire. Specifically, it is responsible for the emergence of a triplet pairing term in addition to the usual induced singlet pairing term. As a result, the system indeed complies with the requirements of Sec.~\ref{sec:RG}, namely it would be driven by repulsive interactions to the TRITOPS phase.

In principle, to quantitatively account for the effect of the spin-orbit coupling term, one needs to have knowledge of the functional form of $\bs\lambda(y,z)$. Deriving $\bs\lambda(y,z)$ from a microscopic theory, however, is a formidable task which we do not attempt here. Instead we shall rely on symmetry considerations, while treating $\bs\lambda(y,z)$ perturbatively, in order to infer its main effect on the low-energy theory.

To construct a tunneling Hamiltonian we introduce an insulating layer between the SC and the nanowire. The width of the layer is $w_b$ and the hight of the potential barrier is $V_b$. The nanowire occupies the space defined by $y\in[-w_y/2,w_y/2]$, $z\in[-w_z/2,w_z/2]$, and is infinite along the $x$ direction. The SC occupies the half space defined by $y<-(w_\tx{b}+w_y/2)$ as depicted in Fig.~\ref{fig:NW_and_conf_pot}.

Following Bardeen~\cite{Bardeen1961tunneling}, we solve for the eigenfunctions in the nanowire $\phi_{k_x}(\boldsymbol{r})$ of the Hamiltonian $\mathcal{H}_{\rm sm}$ but with the potential barrier extended to $y\to-\infty$, and for the eigenfunctions in the normal state of the SC $\chi_{\bf k}(\boldsymbol{r})$ with the potential barrier extended to $y\to\infty$. The tunneling matrix elements are then given by
\begin{equation}
T_{\boldsymbol{k},k_x'}=\int \tx{d}^3\boldsymbol{r}\,\chi^\ast_{\boldsymbol{k}}\left[\mathcal{H}-E_{k_x'}\right]\phi_{k_x'},
\label{eq:Bardeen_mat_elem}
\end{equation}
where $E_{k_x'}$ is the corresponding eigenenergy of $\phi_{k_x'}$, and $\mathcal{H}$ is the Hamiltonian with the true confining potential as depicted in Fig.~\ref{fig:NW_and_conf_pot}.

We solve $\mathcal{H}_{\rm sm}$ in the limit of a high barrier, $\eta_\tx{b}\equiv1/\sqrt{2m_{\rm sm}V_\tx{b}}w_y\ll1$, and we concentrate on energies much smaller than $V_\tx{b}$. To first order in $\eta_\tx{b}$, and to zeroth order in $\bs\lambda(y,z)$ one has
\begin{equation}
\begin{split}
&\phi^{(0)}_{m,n,k_x}(\boldsymbol{r})=\sqrt{\frac{2}{\pi w_yw_z}}e^{ik_xx}\sin[\frac{\pi m}{w_z}(z+\frac{w_z}{2})]\\
&\times \left\{\begin{matrix}\sin[\frac{\pi(1-\eta_\tx{b}) n}{w_y}(y+\frac{w_y}{2})],&-\frac{w_y}{2}< y\le\frac{w_y}{2}\\(-1)^n\pi n\eta_\tx{b} e^{\gamma_\tx{b}(y+w_y/2)},&y\le-\frac{w_y}{2}\end{matrix}\right.,
\end{split}
\end{equation}
with $\gamma_\tx{b}\equiv\sqrt{2m_{\rm sm}V_\tx{b}}$, and where $m,n\in\mathbb{N}$. The eigenenergies are
\begin{equation}
E^{(0)}_{m,n,k_x} = \frac{k_x^2}{2m_{\rm sm}}+\frac{(\pi m)^2}{2m_{\rm sm}w_z^2}+\frac{(\pi n)^2}{2m_{\rm sm}w_y^2}(1-\eta_\tx{b}).
\end{equation}
The eigenfunctions of the SC in the normal state are
\begin{equation}
\begin{split}
&\chi_{\bf k}=\frac{1}{\sqrt{2\pi^3}}e^{i(k_xx+k_zz)}\times\\
&\left\{\begin{matrix} e^{ik_y(y+\frac{w_y}{2}+w_\tx{b})}+\frac{ik_y+\gamma_\tx{b}}{ik_y-\gamma_\tx{b}}e^{-ik_y(y+\frac{w_y}{2}+w_\tx{b})} ,&y<-\frac{w_y}{2}\\ \frac{2ik_y}{ik_y-\gamma_\tx{b}}e^{-\gamma_\tx{b}(y+\frac{w_y}{2}+w_\tx{b})} ,&y\ge-\frac{w_y}{2}\end{matrix}\right..
\end{split}
\end{equation}

We now turn to the first-order corrections of both the energies and the wave functions in the nanowire due to spin-orbit coupling. From symmetry considerations we can infer that $\lambda_x=0$. To see this we first note that the vector field $\bs\lambda$ stems from the electric field in the wire. Since the system is translationally invariant and symmetric under mirror reflection $x\to-x$, the field component $\lambda_x$ must be zero. Moreover, since the system is symmetric under $z\to-z$, we must have $\lambda_z(y,-z)=-\lambda_z(y,z)$. Taking into account the fact that the wave functions $\phi^{(0)}_{m,n,k_x}$ have a definite parity under $z\to-z$, the first-order correction to the energies is given by
\begin{equation}
\begin{split}
E^{(1)}_{m,n,k_x,s} =& \langle\phi^{(0)}_{m,n,k_x}|-i\bs\lambda\cdot(\boldsymbol{\sigma}_{ss}\times\nabla)|\phi^{(0)}_{m,n,k_x}\rangle\\
=&\langle\phi^{(0)}_{m,n,k_x}|\lambda_y|\phi^{(0)}_{m,n,k_x}\rangle k_xs\equiv\alpha k_xs.
\end{split}
\label{eq:Energ_1st_correc}
\end{equation}
where $s=1$ for spin $\uparrow$, and $s=-1$ for spin $\downarrow$. We note that this term vanishes for a system with a symmetry $y\to-y$. It is the breaking of this symmetry by the SC which allows for a nonzero $\alpha$. This is the usual term considered in one-dimensional Rashba systems~\cite{Oreg2010helical,Lutchyn2010majorana}.

We now wish to obtain a correction to the wave functions. We concentrate on the lowest transverse band, namely $m,n=1$, which is justified for a thin wire. We make use of the limit $w_z\ll w_y$, and accordingly consider only the correction due to the second lowest transverse band $|\phi_{1,2,k_x}\rangle$,
\begin{equation}
|\phi^{(1)}_{1,1,k_x}\rangle=\frac{\langle\phi^{(0)}_{1,2,k_x}|-i\bs\lambda\cdot(\boldsymbol{\sigma}\times\nabla)|\phi^{(0)}_{1,1,k_x}\rangle}{E^{(0)}_{1,1,k_x}-E^{(0)}_{1,2,k_x}}|\phi^{(0)}_{1,2,k_x}\rangle.
\end{equation}
Invoking once more the symmetry $\lambda_z(y,-z)=-\lambda_z(y,z)$, one obtains to first order
\begin{equation}
|\phi_{1,1,k_x}\rangle = |\phi^{(0)}_{1,1,k_x}\rangle + \frac{1}{2}\beta k_x\sigma^z|\phi^{(0)}_{1,2,k_x}\rangle,
\label{eq:WF_1st_correc}
\end{equation}
where for the sake of brevity we have defined
\begin{equation}
\beta=\frac{8m_{\rm sm}w_y^2\langle\phi^{(0)}_{1,1,k_x}|\lambda_y|\phi^{(0)}_{1,2,k_x}\rangle}{3\pi^2} .
\end{equation}
This term survives even if the system is symmetric under $y\to-y$, i.e. its existence does not rely on a substrate which breaks inversion symmetry. Its main effect is to push the wave functions either towards or away from the SC, depending on the sign of $k_x\sigma^z$~\cite{Moroz1999Effect}.

We now plug Eq.~\eqref{eq:WF_1st_correc} and Eq.~\eqref{eq:Energ_1st_correc} into Eq.~\eqref{eq:Bardeen_mat_elem} to obtain the matrix elements between modes in the SC and modes in the nanowire. We invoke the limit of a high barrier in which the energies of all the modes are smaller than $V_\tx{b}$, and further assume $k_zw_z\ll1$. This yields
\begin{equation}
\begin{split}
&T_{\boldsymbol{k},k_x'}=t_{\boldsymbol{k}}\delta(k_x-k_x'),\\
&t_{\boldsymbol{k}}=t_0\cos\Theta_{\boldsymbol{k}}(1+\frac{1}{2}\beta k_x'\sigma^z),
\end{split}
\end{equation}
with
\begin{equation}
t_0=\frac{4i|k|}{m_{\rm sm}^2w_yV_\tx{b}}\sqrt{\frac{w_z}{w_y}}e^{-\gamma_\tx{b}w_b},
\end{equation}
and with $\cos\Theta_{\boldsymbol{k}}\equiv k_y/|k|$. Apparently the effect of the inversion-symmetric part of $\lambda_y$ (which is the source of $\beta$) is to introduce a term $k_x\sigma^z$ in the coupling between the wire and the SC. The presence of the factor $\cos\Theta_{\boldsymbol{k}}$ stems simply from the fact that modes which approach the surface of the SC at small angles have a higher probability to tunnel into the wire.

We can now write the full tunneling Hamiltonian of the system as
\begin{equation}
\begin{split}
&H = H_{\rm sm}+H_{\rm sc}+H_{\rm T},\\
&H_{\rm sm} = \frac{1}{2}\int \tx{d}k_x\Phi^\dag_{k_x}\mathcal{H}^{\mathsmaller{\rm BdG}}_{\rm sm}(k_x)\Phi^{\phantom{\dag}}_{k_x},\\
&H_{\rm sc} = \frac{1}{2}\int \tx{d}^3\boldsymbol{k}\Psi^\dag_{\boldsymbol{k}}\mathcal{H}^{\mathsmaller{\rm BdG}}_{\rm sc}(\boldsymbol{k})\Psi^{\phantom{\dag}}_{\boldsymbol{k}},\\
&H_{\rm T} = \frac{1}{2}\int \tx{d}^3{\boldsymbol{k}}\,t_{\boldsymbol{k}}\Psi^\dag_{\boldsymbol{k},s}\Phi^{\phantom{\dag}}_{k_x},\\
&\mathcal{H}^{\mathsmaller{\rm BdG}}_{\rm sm}(k_x)= (\eps_{k_x}-\mu_{\rm sm}+\alpha k_x\sigma^z)\tau^z ,\\
&\mathcal{H}^{\mathsmaller{\rm BdG}}_{\rm sc}(\boldsymbol{k})= \xi_{\boldsymbol{k}}\tau^z+\Delta_{\rm sc}\tau^x ,
\end{split}
\label{eq:full_H_BdG}
\end{equation}
where $\xi_{\boldsymbol{k}}=\boldsymbol{k}^2/2m_{\rm sc}-\mu_{\rm sc}$, $\eps_{k_x}=k_x^2/2m_{\rm sm}-\mu_{\rm sm}$, and with $\Phi_{k_x'}=(c^\dag_{k_x\uparrow},c^\dag_{k_x\downarrow},c^{\phantom\dag}_{-k_x\downarrow},-c^{\phantom\dag}_{-k_x\uparrow})$, $\Psi_{\boldsymbol{k}}=(f^\dag_{\boldsymbol{k}\uparrow},f^\dag_{\boldsymbol{k}\downarrow},f^{\phantom\dag}_{-\boldsymbol{k}\downarrow},-f^{\phantom\dag}_{-\boldsymbol{k}\uparrow})$. Here, $c^\dag_{k_xs}$ create a spin-$s$ electron in the state $\phi_{1,1,k_x}$ of the wire, and $f^\dag_{\boldsymbol{k}s}$ creates a spin-$s$ electron in the state $\chi_{\boldsymbol{k}}$ of the SC. $\{\tau^i\}_{i=x,y,z}$ is a set of Pauli matrices in particle-hole space.

To obtain an effective Hamiltonian for the wire we integrate out the supeconductor's degrees of freedom.~\cite{Sau2010robustness,Stanescu2011majorana,Alicea2012,Danon2015interaction}. The self-energy term which adds to $\mathcal{H}^{\mathsmaller{\rm BdG}}_{\rm sm}(k_x)$ is given by
\begin{equation}
\Sigma(\omega,k_x)=\int \tx{d}k_y\tx{d}k_z \,t_{\boldsymbol{k}}G_{\rm sc}(\omega,\boldsymbol{k})t^\ast_{\boldsymbol{k}},\\
\label{eq:integ_out}
\end{equation}
where $G_{\rm sc}(\omega,\boldsymbol{k})$ is the Green function of the bare SC, given by
\begin{equation}
G_{\rm sc}(\omega,\boldsymbol{k}) = \frac{\omega+\xi_{\boldsymbol{k}}\tau^z-\Delta_{\rm sc}\tau^x}{\omega^2-\xi^2_{\boldsymbol{k}}-\Delta_{\rm sc}^2}.
\end{equation}
To perform the integral in ~\eqref{eq:integ_out} we use the fact that $\mu_{\rm sc}$ is typically much bigger than the relevant energy scale in the semiconductor wire, so we can neglect $k_x^2/2m_{\rm sc}$ compared to $\mu_{\rm sc}$. For the same reason we also have $\mu_{\rm sc}\gg \omega$ which means that the main contribution to the integral comes from momenta satisfying $(k_y^2+k_z^2)/2m_{\rm sc}\simeq\mu_{\rm sc}$. With the help of these simplifications one obtains to first order in $\beta$
\begin{equation}
\Sigma(\omega,k_x)=\frac{\nu_{2d}|t_0|^2(-\omega+\Delta_{\rm sc}\tau^x)}{\sqrt{\Delta_{\rm sc}^2-\omega^2}}(1+\beta k_x\sigma^z),
\end{equation}
where $\nu_{2d}$
is the density of states of a two-dimensional system with an effective mass $m_{\rm sc}$ at a chemical potential $\mu_{\rm sc}$.

Finally, in case one concentrates on energies much smaller than the bare superconducting gap (namely $\omega\ll\Delta_{\rm sc}$), the self-energy becomes independent of $\omega$ and the effective low-energy Hamiltonian is given by
\begin{equation}\label{eq:H_sm_eff_BDG}
\mathcal{H}_{\rm sm}^{\rm eff}=(\eps_{k_x}+\alpha k_x\sigma^z)\tau^z+\Delta_{\rm ind}(1+\beta k_x\sigma^z)\tau^x,
\end{equation}
with $\Delta_{\rm ind}=\nu_{2d}|t_0|^2$.

\section{Derivation of RG flow equations}
\label{sec:RG_eqs_deriv}

In this section we derive the flow equations of Eq.~\eqref{eq:flow_eqs} using a perturbative RG procedure. The action corresponding to the full Hamiltonian $H_0+H_\Delta+H_\tx{int}$, specified in Eqs.~\eqref{eq:minimal_H} and~\eqref{eq:H_int}, is given by $S=S_0 + S_\Delta + S_\tx{int}$, with
\begin{equation}
\begin{split}
&S_0 = -\sum_s\int_{k,\om}\hskip -3mm \left[ (G^\tx{R}_{k\om s})^{-1} \bar{R}_{k\om s} R_{k\om s} +
(G^\tx{L}_{k \om s})^{-1} \bar{L}_{k\om s} L_{k\om s}\right],\\
&S_\Delta = \sum_{s_1s_2} \Delta_{s_1s_2} \int_{k,\om} \hskip -3mm \left( \bar{R}_{k\om s_1}\bar{L}_{-k-\om s_2} +
L_{-k-\om s_2}R_{k\om s_1}\right),
\\
&S_\tx{int}= \int_{1234}u^{s_1s_2}_{s_3s_4}\bar{R}_{k_1\om_1s_1}\bar{L}_{k_2\om_2s_2}L_{k_3\om_3s_3}R_{k_4\om_4s_4},\\
\end{split}
\end{equation}
where $R_{k\om s}$, $\bar{R}_{k\om s}$, $L_{k\om s}$, and $\bar{L}_{k\om s}$ are Grassman fields, and where we have used the abbreviations
\begin{equation}
\int_{k,\om} \equiv \int_{-\infty}^\infty \frac{\tx{d}\om}{2\pi}\int_{-\Lambda}^\Lambda \frac{\tx{d}k}{2\pi},
\end{equation}
and
\begin{equation}\label{eq:int1234}
\begin{split}
\int_{1234} \equiv & (2\pi)^2 \prod_{i=1}^4 \sum_{s_i}
\int_{-\Lambda}^\Lambda\frac{\tx{d}k_i}{2\pi}
\int_{-\infty}^\infty\frac{\tx{d}\om_i}{2\pi}\times\\
&\delta(k_1+k_2-k_3-k_4) \delta(\om_1+\om_2-\om_3-\om_4).
\end{split}
\end{equation}
Above we have used a compact notation for the action $S$, by using the Green functions $G^\tx{R,L}_{k\om s}$ and the couplings $\Delta_{s_1s_2}$ and $u^{s_1s_2}_{s_3s_4}$, which are defined by
\begin{subequations}
\begin{align}
\label{eq:Green_funcs}
G^\eta_{k\om s} =& \left(i\om-\eta \cdot v_{\eta\cdot s}k \right)^{-1},
\\
\label{eq:Delta12}
\Delta_{s_1s_2} = & \Delta_\tx{s} i \sigma^y_{s_1s_2} + \Delta_\tx{t} \sigma^x_{s_1s_2},
\\
\label{eq:u1234}
\begin{split}
u^{s_1s_2}_{s_3s_4}=
&-g_1^\perp\sigma^x_{s_1s_2}\sigma^x_{s_2s_3}\sigma^x_{s_3s_4}
+g_2^\parallel\delta_{s_1s_2}\delta_{s_2s_3}\delta_{s_3s_4}\\
&+(g_2^+\delta_{s_1\ua}+g_2^-\delta_{s_1\da})\sigma^x_{s_1s_2}\delta_{s_2s_3}\sigma^x_{s_3s_4}.
\end{split}
\end{align}
\end{subequations}
On the right-hand side of Eq.~\eqref{eq:Green_funcs} we have used a convention where $\eta=\tx{R}(\tx{L})$ corresponds to $\eta=1(-1)$, and $s=\ua(\da)$ corresponds to $s=1(-1)$.

To study the low-energy physics of the system, we iteratively integrate out the high-momentum modes within a small momentum shell, thereby obtaining an action with an effectively-decreasing cutoff, $\Lambda\exp(-\ell)$, where $\ell$ is the so-called RG time~\cite{Shankar1994Renormalization}. We are interested in the flow of the couplings $\Delta_+$, $\Delta_+$, $g_1^\perp$, $g_2^\parallel$, $g_2^+$, and $g_2^-$ as a function of $\ell$.

At tree level, all the interaction couplings $g_1^\perp$, $g_2^\parallel$, $g_2^+$, and $g_2^-$ are marginal with respect to the fixed point action $S_0$. The induced pairing potentials $\Delta_\tx{s,t}$ (or equivalently $\Delta_\pm$) are relevant, on the other hand, with a scaling dimension of $1$. Importantly, the one-loop corrections will cause a difference in the flow of $\Delta_\tx{s}$ and $\Delta_\tx{t}$.

\begin{figure}
\begin{tabular}{lr}
\rule{0pt}{15ex}
\includegraphics[clip=true,trim =0cm 0cm 0cm 0cm,scale=0.35]{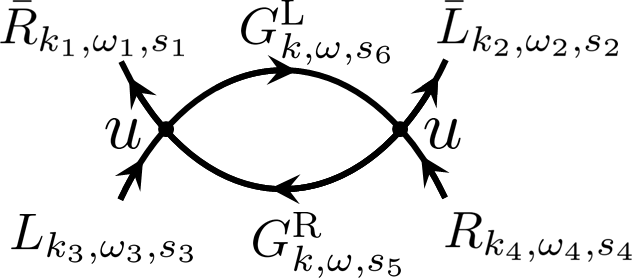}
\llap{\parbox[c]{7.9cm}{\vspace{-35mm}\footnotesize{(a)}}} & \hskip 1mm
\includegraphics[clip=true,trim =0cm 0cm 0cm 0cm,scale=0.35]{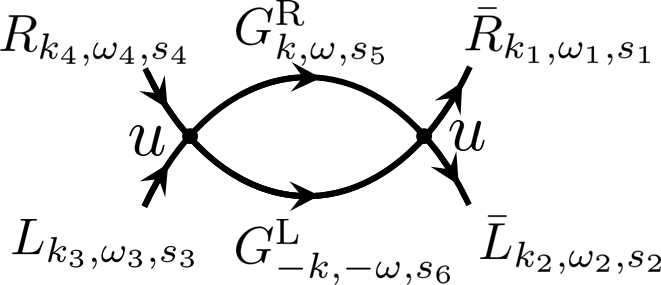}
\llap{\parbox[c]{8.2cm}{\vspace{-35mm}\footnotesize{(b)}}} \\ \rule{0pt}{12ex}
\includegraphics[clip=true,trim =0cm 0cm 0cm 0cm,scale=0.35]{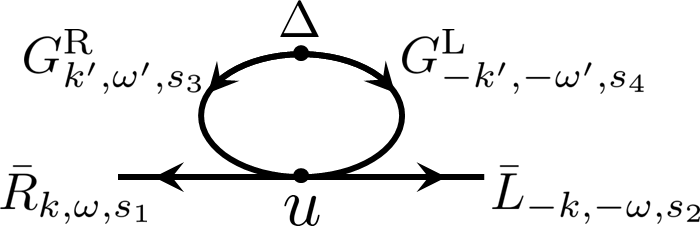}
\llap{\parbox[c]{8.6cm}{\vspace{-27mm}\footnotesize{(c)}}}
\end{tabular}
\caption{Diagrammatic representation of the second-order corrections to the $S_0$, appearing in Eq.~\eqref{eq:one_loop_correc}. In (a) and (b) are diagrams which renormalize the interaction couplings, while in (c) is the diagram which renormalizes the induced pairing potentials. The corrections due to these diagrams are denoted by $\delta u^\tx{ZS}$ and $\delta u^\tx{BCS}$, and $\delta\Delta$,  respectively.}\label{fig:Diagrams}
\end{figure}

To obtain the one-loop corrections to the flow, we treat $S'=S_\Delta + S_\tx{int}$ as a perturbation to $S_0$ and apply the cumulant expansion. Integrating over the fast modes, one has
\begin{equation}
\delta S=\frac{1}{2}\left(\langle S'\rangle_{0,>}^2-\langle S'^2\rangle_{0,>}\right),
\end{equation}
where $\langle\hskip 2mm\rangle_{0,>}$ stands for averaging over the fast modes with respect to the unperturbed action $S_0$. This results in the following corrections
\begin{subequations}\label{eq:one_loop_correc}
\begin{align}
\label{eq:d_u_BCS}
(&\delta u^\tx{BSC})^{s_1s_2}_{s_3s_4} = -
\sum_{s_5s_6}u^{s_1s_2}_{s_6s_5}u^{s_5s_6}_{s_3s_4}\int_{k^>,\om} G^\tx{R}_{k,\om,s_5}G^\tx{L}_{-k,-\om,s_6},\\
\label{eq:d_u_ZS}
(&\delta u^\tx{ZS})^{s_1s_2}_{s_3s_4} = -
\sum_{s_5s_6}u^{s_1s_6}_{s_3s_5}u^{s_5s_2}_{s_6s_4}\int_{k^>,\om} G^\tx{R}_{k,\om,s_5}G^\tx{L}_{k,\om,s_6},\\
\label{eq:d_Delta}
&\delta \Delta_{s_1s_2} =
-\sum_{s_3s_4}u^{s_1s_2}_{s_4s_3}\Delta_{s_3s_4}\int_{k^>,\om}G^\tx{R}_{k,\om,s_3}G^\tx{L}_{-k,-\om,s_4},
\end{align}
\end{subequations}
which are described diagrammatically in Fig.~\ref{fig:Diagrams}. In obtaining Eqs.~(\ref{eq:d_u_BCS},\ref{eq:d_u_ZS}) we have set the momenta and frequencies of the outer (slow) legs [see Fig.~\hyperref[fig:Diagrams]{\ref{fig:Diagrams}(a,b)}] to zero ~\cite{Shankar1994Renormalization}.

Finally, one can perform the frequency and momentum integration in Eq.~\eqref{eq:one_loop_correc} and arrive at
\begin{subequations}\label{eq:flow_eqs_raw}
\begin{align}
&\dot{g}_1^\perp = \frac{1}{\pi}\left( \frac{2}{v_++v_-}g_2^\parallel - \frac{1}{2v_+}g_2^+ - \frac{1}{2v_-}g_2^- \right)g_1^\perp ,
\\
&\dot{g}_2^\parallel = \frac{1}{\pi(v_++v_-)} {g_1^\perp}^2 ,
\\
&\dot{g}_2^+ = -\frac{1}{2\pi v_-} {g_1^\perp}^2 ,
\\
&\dot{g}_2^- = -\frac{1}{2\pi v_+} {g_1^\perp}^2 ,
\\
&\dot{\Delta}_+ = \Delta_+ - \frac{1}{2\pi v_+} g_2^+\Delta_+ - \frac{1}{2\pi v_-} g_1^\perp \Delta_- ,
\\
&\dot{\Delta}_- = \Delta_- - \frac{1}{2\pi v_-} g_2^-\Delta_- - \frac{1}{2\pi v_+} g_1^\perp \Delta_+ .
\end{align}
\end{subequations}
Defining the average velocity, $\bar{v}=(v_++v_-)/2$, and the dimensionless couplings, $y_1^\perp=g_1^\perp/\pi\bar{v}$, $y_2^+=g_2^+/\pi v_+$, $y_2^-=g_2^-/\pi v_-$, and $y_2=g_2^+/2\pi v_+ + g_2^-/2\pi v_- - g_2^\parallel/\pi\bar{v}$, one immediately arrives at Eq.~\eqref{eq:flow_eqs}.

As noted in the Sec.~\ref{sec:RG}, we solve the flow equations up to an RG time $\ell^\ast$, defined as the time at which one of the pairing potentials flows to strong coupling (meaning it becomes of the order of the energy cutoff). Let us assume, for example, that $\Delta_+$ flows to strong coupling first, namely that $|\Delta_+(\ell^\ast)|=v_+\Lambda$. The positive-helicity degrees of freedom, $R_{k\om\ua}$ and $L_{k\om\da}$ are therefore gapped and we can integrate them out. We are then left with an action containing only the negative-helicity fields $R_{k\om\da}$ and $L_{k\om\ua}$,
\begin{equation}\label{eq:S_minus}
\begin{split}
S_- =& \int_{k,\om}\hskip -3mm\left\{ -\left[ (G^\tx{R}_{k\om \da})^{-1} \bar{R}_{k\om \da} R_{k\om \da} +
(G^\tx{L}_{k \om \ua})^{-1} \bar{L}_{k\om \ua} L_{k\om \ua}\right]\right.\\
&\left.+\Delta_-'(\ell^\ast) \left( \bar{L}_{k\om \ua}\bar{R}_{-k-\om \da} +
R_{-k-\om \da}L_{k\om \ua}\right)\right\}\\
+&g_2^-(\ell^\ast)\int_{1234} \bar{R}_{k_1\om_1\da} \bar{L}_{k_2\om_2\ua} \bar{L}_{k_3\om_3\ua} R_{k_4\om_4\da},
\end{split}
\end{equation}
where to leading order in the interaction couplings
\begin{equation}\label{Delta_minus_prime}
\begin{split}
\Delta_-'(\ell^\ast) &= \Delta_-(\ell^\ast) + g_1^\perp(\ell^\ast) \int_{k\om}\int_{k',\om'}\langle L_{k\om\da} R_{k'\om'\ua} \rangle_+ = \\
&= \Delta_-(\ell^\ast) - \frac{g_1^\perp(\ell^\ast)}{2\pi v_+}\Delta_+(\ell^\ast) \sinh\left[\frac{v_+\Lambda}{|\Delta_+(\ell^\ast)|}\right],
\end{split}
\end{equation}
and where $\avg{\hskip 1mm}_+$ stands for averaging with respect to the action containing only the positive-helicity fields. We can now continue with the RG procedure, applied to $S_-$, which results in the following flow equations
\begin{subequations}
\begin{align}\label{eq:flow_eqs_scd_step_appn}
\dot{g}_2^- &= 0,\\
\dot{\Delta}'_- &= \left(1 - \frac{g_2^-}{2\pi v_-}\right)\Delta'_-.
\end{align}
\end{subequations}
The flow is again stopped when $\Delta'_-$ reaches strong coupling. Importantly, the sign of the gap is determined by the sign of $\Delta_-'(\ell^\ast)$. The topological invariant is then given by $\mc{Q}=\sgn[\Delta_+(\ell^\ast)]\sgn[\Delta_-'(\ell^\ast)]$.

Finally, let us consider the possible interaction terms which were not included in Eq.~\eqref{eq:H_int}. To this end, we first turn back attention to Eq.~\eqref{eq:one_loop_correc}. We note that since the frequency integrals of Eq.~\eqref{eq:one_loop_correc} contain one right-moving green-function and one left-moving Green function, there exists poles in both the lower and upper halves of the complex frequency plane. Had the two Green functions been of the same chirality, the two poles would have been in the same half plane, resulting in a vanishing integral. We can now easily consider additional interaction terms which are also allowed by time-reversal symmetry,
\begin{equation}\label{eq:S_int_p}
\begin{split}
H_\tx{int}' = \int\tx{d}x & \left\{g_4^\perp \left[\rho_{\tx{R}\ua}(x)\rho_{\tx{R}\da}(x) + \rho_{\tx{L}\da}(x)\rho_{\tx{L}\ua}(x)\right] \right. \\
& \hskip -2.3mm + \left. g_4^+ \left[\rho_{\tx{R}\ua}(x)\rho_{\tx{R}\ua}(x) + \rho_{\tx{L}\da}(x)\rho_{\tx{L}\da}(x)\right]\right.\\
& \hskip -2.3mm + \left.g_4^- \left[\rho_{\tx{R}\da}(x)\rho_{\tx{R}\da}(x) + \rho_{\tx{L}\ua}(x)\rho_{\tx{L}\ua}(x)\right]\right\} .
\end{split}
\end{equation}
The couplings $g_4^\perp$, $g_4^+$, and $g_4^-$ are marginal at tree level. Considering the above argument, any one-loop correction involving these couplings will necessarily contain a loop with two Green functions of the same chirality, and would therefore vanish. As a result, these couplings do not affect the flow of $\Delta_\pm$, $g_1^\perp$, $g_2^\pm$, and $g_2^\parallel$, nor do they flow by themselves. This is the reason for not considering $H_\tx{int}'$ to begin with.

\bibliography{Refs_TRITOPS_RG}

\begin{thebibliography}{86}
\bibitem [{Pse()}]{PseudoSpinIndex}%
  \BibitemOpen
  \href@noop {} {}\bibinfo {note} {It should be noted that the identification
  of the index $s=\ua,\da$ as the spin is not crucial. One can instead consider
  any two modes $R_1(x)$, $R_2(x)$ and their time-reversal partners $L_2(x)$,
  $L_1(x)$, respectively.}\BibitemShut {Stop}%
\bibitem [{Mom()}]{MomentumMismachTerm}%
  \BibitemOpen
  \href@noop {} {}\bibinfo {note} {If $k_\tx{F}^+=k_\tx{F}^-$ one can have, for
  example, a term $(\Delta'R_\ua L_\ua + \Delta'^\ast L_\da R_\da +
  \tx{h.c.})$. For $k_\tx{F}^+\neq k_\tx{F}^-$, however, this term would be
  suppressed at low energies due to momentum mismatch.}\BibitemShut {Stop}%
\bibitem [{Top()}]{TopInvWeaPairing}%
  \BibitemOpen
  \href@noop {} {}\bibinfo {note} {This topological invariant was derived in
  the weak-pairing limit. However, for the linearized low-energy model of
  Eq.~\eqref{eq:minimal_H} it is exact.}\BibitemShut {Stop}%
\bibitem [{Int()}]{IntSysTopInvComment}%
  \BibitemOpen
  \href@noop {} {}\bibinfo {note} {In Sec.~\ref{sec:RG} we clarify how to apply
  the topological invariant,
  $Q=\sgn(\Delta_+\Delta_-)=\sgn(\Delta_\tx{s}^2-\Delta_\tx{t}^2)$, to a system
  with interactions [see Eq.~\eqref{eq:top_inv_RG}].}\BibitemShut {Stop}%
\bibitem [{Upp()}]{UpperVsLowerBand}%
  \BibitemOpen
  \href@noop {} {}\bibinfo {note} {Similar results are obtained if one consider
  the case where the chemical potential lies inside the lower
  band.}\BibitemShut {Stop}%
\bibitem [{Wea()}]{WeakPairing}%
  \BibitemOpen
  \href@noop {} {}\bibinfo {note} {Weak pairing here means that the pairing
  potential is small compared to the distance from the Fermi energy to the
  bottom of the band, namely $\mu-|t|\ll|\Delta(k)|$.}\BibitemShut {Stop}%
\bibitem [{Est()}]{EstimateEllStar}%
  \BibitemOpen
  \href@noop {} {}\bibinfo {note} {This estimation is obtained upon neglecting
  the second order terms in Eqs.~(\ref{eq:Delta_p_flow},\ref{eq:Delta_m_flow})
  and integrating them up to $\Delta_\pm(\ell^\ast)=v_\pm\Lambda$.}\BibitemShut
  {Stop}%
\end{thebibliography}
\end{document}